\PassOptionsToPackage{table}{xcolor}
\documentclass[10pt]{wlscirep}
\usepackage[utf8]{inputenc}
\usepackage[T1]{fontenc}
\usepackage{lineno}
\usepackage{multirow}
\usepackage{amssymb}%
\usepackage{pifont}%
\usepackage{mathrsfs}
\usepackage{newtxmath}
\usepackage{tabularx}
\usepackage{algorithm}
\usepackage{algorithmic}
\usepackage{bbm}

\definecolor{myred_2}{rgb}{0.99, 0, 0}
\definecolor{myblue_2}{rgb}{0, 0, 0.99}
\definecolor{myblue}{rgb}{0, 0, 0}
\definecolor{red}{rgb}{0, 0, 0}
\definecolor{myred_2}{rgb}{0, 0, 0}
\definecolor{myblue_2}{rgb}{0, 0, 0}

\DeclareMathAlphabet{\mathcal}{OMS}{cmsy}{m}{n}
\newcommand{\cmark}{\ding{51}}%
\newcommand{\xmark}{\ding{55}}%

\newcommand{\myrev}[1]{{\color{myblue}#1}}

\usepackage{shellesc} %

\newif\ifcountwords
\countwordstrue

\title{Mask-prior-guided denoising diffusion improves inverse protein folding}

\author[1,2]{Peizhen Bai}
\author[3]{Filip Miljković}
\author[1,4]{Xianyuan Liu}
\author[5]{Leonardo De Maria}
\author[2]{Rebecca Croasdale-Wood}
\author[6]{Owen Rackham}
\author[1,4,$\dagger$]{Haiping Lu}
\affil[1]{School of Computer Science, University of Sheffield, Sheffield, United Kingdom}
\affil[2]{Biologics Engineering, Oncology R\&D, AstraZeneca, Cambridge, United Kingdom}
\affil[3]{Medicinal Chemistry, Research and Early Development, Cardiovascular, Renal and Metabolism, BioPharmaceuticals R\&D, AstraZeneca, Gothenburg, Sweden}
\affil[4]{Centre for Machine Intelligence, University of Sheffield, Sheffield, United Kingdom}
\affil[5]{Medicinal Chemistry, Research and Early Development, Respiratory and Immunology, BioPharmaceuticals R\&D, AstraZeneca, Gothenburg, Sweden}
\affil[6]{School of Biological Sciences, University of Southampton, Southampton, United Kingdom}

\affil[$\dagger$]{Corresponding author: Haiping Lu (h.lu@sheffield.ac.uk)}

\begin{abstract}
Inverse protein folding generates valid amino acid sequences that can fold into a desired protein structure, with recent deep-learning advances showing strong potential and competitive performance. \textcolor{myblue}{However, challenges remain, such as predicting elements with high structural uncertainty, including disordered regions.} To tackle such low-confidence residue prediction, we propose a \textbf{Ma}sk-\textbf{p}rior-guided denoising \textbf{Diff}usion (\textbf{MapDiff}) framework that accurately captures both structural information and residue interactions for inverse protein folding. MapDiff is a discrete diffusion probabilistic model that iteratively generates amino acid sequences with reduced noise, conditioned on a given protein backbone. To incorporate structural information and residue interactions, we develop a graph-based denoising network with a mask-prior pre-training strategy. Moreover, in the generative process, we combine the denoising diffusion implicit model with Monte-Carlo dropout to \textcolor{myblue}{reduce uncertainty}. Evaluation on four challenging sequence design benchmarks shows that MapDiff substantially outperforms state-of-the-art methods. Furthermore, the \textit{in silico} sequences generated by MapDiff closely resemble the physico-chemical and structural characteristics of native proteins across different protein families and architectures.
\end{abstract}

\begin{document}
\flushbottom
\maketitle
\thispagestyle{empty}

\section*{Introduction}
Proteins are complex, three-dimensional (3D) structures folded from linear amino acid (AA) sequences. They play a critical role in essentially all biological processes, including metabolism, immune response and cell cycle control. The inverse protein folding (IPF) problem is a fundamental structure-based protein design problem in computational biology and medicine. It aims to generate valid AA sequences with the potential to fold into a desired 3D backbone structure, enabling the creation of novel proteins with specific functions \cite{dauparas2022robust}. Its enormous applications range from therapeutic protein engineering, lead compound optimization and antibody design \cite{antifold}. 

Traditional physics-based approaches consider IPF as an energy optimization problem \cite{alford2017rosetta}, suffering from high computational cost and limited accuracy. In recent years, deep learning has emerged as the preferred paradigm for solving protein structure problems due to its strong ability to learn complex non-linear patterns from data adaptively. In deep learning for IPF, early convolutional neural network-based models view each protein residue as an isolated unit or the whole as point cloud data, with limited consideration of structural information and interactions between residues \cite{wu2021protein, li2014direct, o2018spin2, anand2022protein}. Recently, graph-based methods have represented 3D protein structures as proximity graphs, and then use graph neural networks (GNNs) to model residue representations and incorporate structural constraints. GNNs can aggregate and exchange local information within graph-structured data, enabling substantial performance improvement in graph-based methods.

Despite the advances in graph-based methods, structural information alone cannot determine \textcolor{myblue}{the residue identities of some challenging structural elements, such as intrinsically disordered regions \cite{towse2012domain}}. In such uncertain, low-confidence cases, interactions with other accurately predicted residues can provide more reliable guidance for mitigating uncertainty in these regions. Moreover, existing deep learning-based IPF methods typically employ autoregressive decoding or uniformly random decoding to generate AA sequences, prone to accumulate prediction errors \cite{li2020multitask, martinez2021pose} and limited in capturing global and long-range dependencies in protein evolution \cite{starr2016epistasis, xu2021anytime}. Recently, several non-autoregressive alternatives have shown the potential to outperform the autoregressive paradigm in related contexts \cite{gao2023pifold, li2020multitask, lyu2024variational}. Additionally, protein structure prediction methods, such as the AlphaFold series \cite{jumper2021highly, abramson2024accurate}, often take an iterative generation process to refine non-deterministic structures by integrating well-predicted information. These raise the question: can combining residue interactions with an iterative refinement and an efficient non-autoregressive decoding improve IPF prediction performance to generate more plausible protein sequences?

\begin{figure*}[!t]
    \begin{center}
    \includegraphics[width=0.96\textwidth]{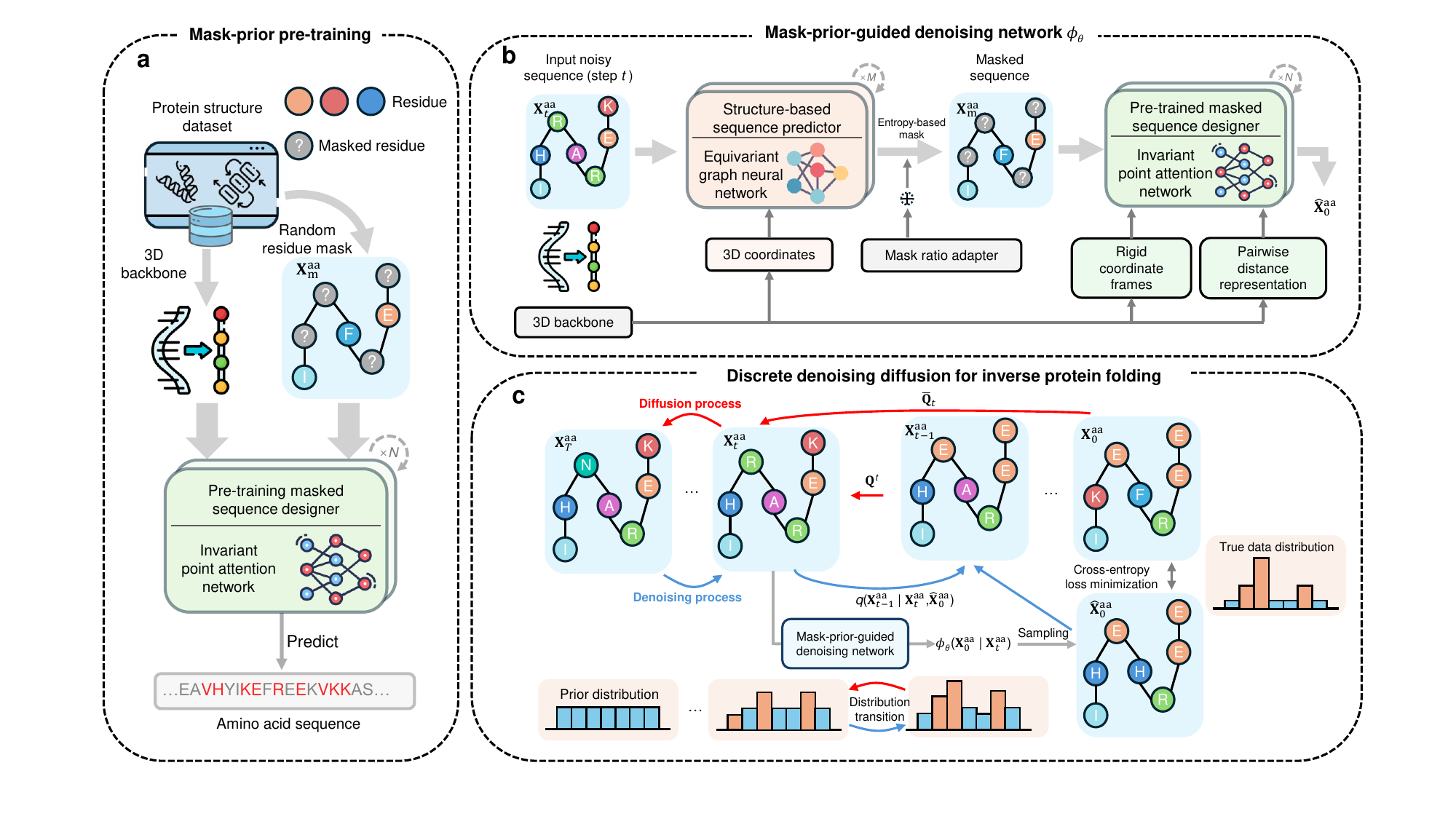}
    \end{center}
    \caption{\textbf{Mask-prior-guided denoising diffusion (MapDiff) for inverse protein folding.} \textbf{(a)} The mask-prior pre-training stage randomly masks residues within the AA sequence and pre-trains an invariant point attention (IPA) network with the masked sequence and the 3D backbone structure to learn prior structural and sequence knowledge, using BERT-like masked language modelling objectives. \textbf{(b)} The mask-prior-guided denoising network $\phi_{\theta}$ takes an input noisy AA sequence $\mathbf{X}^\mathrm{aa}$ to predict the native AA sequence $\mathbf{X}^\mathrm{aa}_0$ via three operations in every iterative denoising step: it first initializes a structure-based sequence predictor as an equivariant graph neural network to denoise the noisy sequence $\mathbf{X}^\mathrm{aa}$ conditioned on the provided 3D backbone structure. Then, combining an entropy-based mask strategy with a mask ratio adaptor identifies and masks low-confidence residues in the denoised sequence in the first step to produce a masked sequence $\mathbf{X}^\mathrm{aa}_\mathrm{m}$. Next, the pre-trained masked sequence designer in (a) takes the masked sequence $\mathbf{X}^\mathrm{aa}_\mathrm{m}$ and its 3D backbone information for refinement (fine-tuning) to better predict the native sequence $\mathbf{X}^\mathrm{aa}_0$. \textbf{(c)} The MapDiff denoising diffusion framework iteratively alternates between two processes: diffusion and denoising. \textcolor{myblue}{The diffusion process progressively adds random discrete noise to the native sequence $\mathbf{X}^\mathrm{aa}_0$ according to the cumulative transition matrix $\overline{\mathbf{Q}}_t$ at the diffusion step $t$ so that the real data distribution can gradually transition to a uniform or marginal prior distribution}. The denoising process randomly samples an initial noisy AA sequence $\mathbf{X}^\mathrm{aa}_T$ from the prior distribution and iteratively uses the denoising network $\phi_{\theta}$ in (b) to denoise it, learning to predict the native sequence $\mathbf{X}^\mathrm{aa}_0$ from $\mathbf{X}^\mathrm{aa}_t$ at each denoising step $t$. The prediction $\hat{\mathbf{X}}^\mathrm{aa}_0$ facilitates the computation of the posterior distribution $q({\mathbf{X}_{t-1}^\mathrm{aa} \mid \mathbf{X}_{t}^\mathrm{aa}, \hat{\mathbf{X}}_{0}^\mathrm{aa}})$ for predicting a less noisy sequence $\mathbf{X}^\mathrm{aa}_{t-1}$.}
    \label{fig:architecture}
\end{figure*}

Recently, denoising diffusion models, an innovative class of deep generative models, have gained growing attention in various fields. They learn to generate conditional or unconditional data by iteratively denoising random samples from a prior distribution. Diffusion-based models have been adopted for de novo protein design and molecule generation, achieving state-of-the-art performance. For instance, RFdiffusion \cite{watson2023novo} fine-tunes the protein structure prediction network RoseTTAFold \cite{baek2021accurate} under a denoising diffusion framework to generate 3D protein backbones, and Torsional Diffusion \cite{jing2022torsional} implements a diffusion process on the space of torsion angles for molecular conformer generation. In structure-based drug design, DiffSBDD \cite{schneuing2022structure} proposes an equivariant 3D-conditional diffusion model to generate novel small-molecule binders conditioned on target protein pockets. While diffusion models have a widespread application in computational biology, most existing methods primarily focus on generating structures in continuous 3D space. The potential of diffusion models in inverse folding has not been fully exploited yet.

We propose a \textbf{Ma}sk-\textbf{p}rior-guided denoising \textbf{Diff}usion (\textbf{MapDiff}) framework (Fig. \ref{fig:architecture}) to accurately capture structure-to-sequence mapping for IPF prediction. Unlike previous graph-based methods, MapDiff models IPF as a discrete denoising diffusion problem that iteratively generates less-noisy AA sequences conditioned on a target protein structure. Due to the property of denoising diffusion, MapDiff can also be viewed as an iterative refinement that enhances the accuracy of the generated sequences over time. Moreover, \textcolor{myblue}{we design a two-step denoising network to adaptively improve the denoising trajectories using a pre-trained mask prior.} Our denoising network effectively leverages the structural information and residue interactions to reduce prediction error on low-confidence residue prediction. To further improve the denoising speed and uncertainty estimation, we combine the denoising diffusion implicit model (DDIM) \cite{song2021denoising}  with Monte-Carlo dropout \cite{gal2016dropout} in the discrete generative process. \textcolor{red}{DDIM accelerates sequence generation by skipping multiple denoising steps, while Monte-Carlo dropout reduces uncertainty by performing multiple stochastic forward passes with dropout enabled during inference.} We conduct 
\myrev{performance comparisons}
against state-of-the-art methods for IPF prediction, demonstrating the effectiveness of MapDiff across multiple metrics and benchmarks, outperforming even those incorporating external knowledge. Moreover, when we use AlphaFold2 \cite{jumper2021highly} to fold the sequences generated by MapDiff back to 3D structures, such AlphaFold2-folded structures are highly similar to the native protein templates, even for cases of low sequence recovery rates.

This work shows the high potential of utilizing discrete denoising diffusion models with mask-prior pre-training for IPF prediction. Our main contributions are three-fold: (i) we propose a discrete denoising diffusion-based framework named MapDiff to explicitly consider the structural information and residue interactions in the diffusion and denoising processes; (ii) we design a mask-prior-guided denoising network that adaptively denoise the diffusion trajectories to produce feasible and diverse sequences from a fixed structure; (iii) MapDiff incorporates discrete DDIM with Monte-Carlo dropout to accelerate the generative process and improve uncertainty estimation.

\section*{Results}

\subsection*{MapDiff framework}
As shown in Fig. \ref{fig:architecture}, the MapDiff framework formulates IPF prediction as a denoising diffusion problem (Fig. \ref{fig:architecture}c). The diffusion process progressively adds random discrete noise to the native AA sequence according to the transition probability matrices to facilitate the training of a denoising network. In the denoising process, this denoising network iteratively denoises a noisy, randomly sampled AA sequence conditioned on the 3D structural information to predict or reconstruct the native AA sequence. The diffusion and denoising processes iterate alternately to capture the sampling diversity of native sequences from their complex distribution and refine the predicted AA sequences.

We propose a novel mask-prior-guided denoising network to adaptively adjust the discrete denoising trajectories towards generating more valid AA sequences via three operations within each iterative denoising step (Fig. \ref{fig:architecture}b). Firstly, a structure-based sequence predictor employs an equivariant graph neural network (EGNN) \cite{satorras2021n} to denoise the noisy sequence conditioned on the backbone structure. Secondly, we use an entropy-based mask strategy \cite{zhou2023prorefiner} and a mask ratio adaptor to identify and mask low-confidence or uncertain (e.g. structurally undetermined) residues in the denoised sequence in the first operation to produce a masked sequence. Thirdly, a pre-trained masked sequence designer network predicts the masked residues to obtain their refined prediction. The pre-training of the masked sequence designer is done before the diffusion and denoising processes via an invariant point attention (IPA) network \cite{jumper2021highly} using masked language modelling (Fig. \ref{fig:architecture}a), incorporating prior structural and sequence knowledge. \textcolor{myblue}{The structure-based sequence predictor and masked sequence designer refine denoising trajectories by leveraging structural information and residue interactions. For efficient sequence generation, the denoising network uses non-autoregressive decoding to generate sequences in a one-shot manner \cite{gao2023pifold}. Additionally, we incorporate DDIM \cite{song2021denoising} to accelerate inference by skipping multiple denoising steps and Monte-Carlo dropout \cite{gal2016dropout} to reduce uncertainty. The Methods section provides more details.}

\begin{table*}[t]
   \centering
   \small
   \caption{Performance comparison on the CATH 4.2 and CATH 4.3 datasets with topology classification split. The results include the perplexity and median recovery rate on the full test set, as well as on short and single-chain subsets. The external knowledge column indicates whether additional training data or protein language models are used. We also quoted partial baseline results from Gao et al. (2022) \cite{gao2023pifold} and Hsu et al. (2022) \cite{hsu2022learning} for comparative analysis, marked with \textsuperscript{$\dagger$}.}
   \label{tab:results_cath}
   \resizebox{1\linewidth}{!}{\begin{tabularx}{\textwidth}{lXcccccccc}
   \toprule
   & \multirow{2}{*}{\bf Models} 
   & \multirow{2}{*}{\bf External}
   & \multirow{2}{*}{\bf Model} 
   & \multicolumn{3}{c}{\bf Perplexity ($\downarrow$)} 
   & \multicolumn{3}{c}{\bf Median recovery rate (\%, $\uparrow$)} \\
   \cmidrule[0.3pt](lr){5-7} \cmidrule[0.3pt](lr){8-10}
   & & \bf knowledge & \bf parameters & Short   & Single-chain  & Full  & Short   & Single-chain   & Full \\
   \midrule
   \multirow{10}{*}{\rotatebox[origin=c]{90}{\bf CATH 4.2}}
   & $^\dagger$StructGNN \cite{ingraham2019generative}               & \xmark & 1.4M & 8.29  & 8.74 & 6.40&   29.44   &28.26& 35.91 \\
   & $^\dagger$GraphTrans \cite{ingraham2019generative}              & \xmark & 1.5M & 8.39  &   8.83 &  6.63 &   28.14 & 28.46 & 35.82 \\
   & $^\dagger$GVP \cite{jing2020learning}              & \xmark & 2.0M & 7.09  &   7.49  &  6.05 &  32.62  & 31.10  & 37.64 \\
   & $^\dagger$AlphaDesign \cite{gao2022alphadesign}            & \cmark & 6.6M & 7.32  &   7.63  &  6.30 &   34.16 & 32.66 &  41.31 \\ 
  \cmidrule[0.3pt](lr){2-10}
  & ProteinMPNN \cite{dauparas2022robust}                    & \xmark &1.9M &6.90 &7.03  & 4.70 & 36.45 & 35.29 & 48.63 \\ 
  & PiFold \cite{gao2023pifold}                 & \xmark &6.6M & 5.97 & \underline{6.13} & 4.61 & 39.17 & 42.43 & 51.40 \\ 
  & LM-Design \cite{lmdesign}        & \cmark &659M & 6.86 & 6.82 & 4.55 & 37.66 & 38.94 & \underline{53.19} \\ 
  & GRADE-IF \cite{yi2024graph}         & \xmark &7.0M & \underline{5.65} & 6.46 &\underline{4.40} & \underline{45.84} & \underline{42.73} & 52.63 \\  
  \cmidrule[0.3pt](lr){2-10}
  & MapDiff (uniform prior)   & \xmark  & 14.7M   & 3.99 & 4.43 & 3.46 & 52.85 & \textbf{50.00} & \textbf{61.03} \\
  & MapDiff (marginal prior)    & \xmark  & 14.7M   & \textbf{3.96} & \textbf{4.41} & \textbf{3.43} & \textbf{54.04} & 49.34 & 60.93 \\
  \midrule
   \multirow{10}{*}{\rotatebox[origin=c]{90}{\bf CATH 4.3}}
  & $^\dagger$GVP-GNN-Large \cite{hsu2022learning}   & \xmark & 21M & 7.68  & 6.12  & 6.17 & 32.60 & 39.40 & 39.20 \\
  & $^\dagger$+ AF2 predicted data    & \cmark & 142M & 6.11  & 4.09  & 4.08 & 38.30 & 50.08 & 50.08 \\
  & $^\dagger$GVP-Transformer \cite{hsu2022learning}    & \xmark & 21M & 8.18  & 6.33  & 6.44 & 31.30 & 38.50 & 38.30 \\
  & $^\dagger$+ AF2 predicted data    & \cmark & 142M & 6.05  & 4.00  & 4.01 & 38.10 & 51.50 & 51.60 \\
  \cmidrule[0.3pt](lr){2-10}
  & ProteinMPNN \cite{dauparas2022robust}              & \xmark  & 1.9M & 6.12 & 6.18 & 4.63 & 40.00 & 39.13 & 47.66 \\ 
  & PiFold \cite{gao2023pifold}            & \xmark & 6.6M & 5.52 & \underline{5.00} & \underline{4.38} & 43.06 & 45.54 & 51.45 \\ 
  & LM-Design \cite{lmdesign}     & \cmark & 659M & 6.01 & 5.73 & 4.47 & 44.44 & 45.31 & \underline{53.66} \\ 
  & GRADE-IF \cite{yi2024graph}       & \xmark & 7.0M & \underline{5.30} & 6.05 & 4.58 & \underline{48.21} & \underline{45.94} & 52.24 \\ 
  \cmidrule[0.3pt](lr){2-10}
  & MapDiff (uniform prior)   & \xmark  & 14.7M   & \textbf{3.88} & 3.85 & \textbf{3.48} & \textbf{55.95} & 54.65 & \textbf{60.86} \\
  & MapDiff (marginal prior)   & \xmark  & 14.7M   & 3.90 & \textbf{3.83} & 3.52 & 55.56 & \textbf{54.99} & 60.68 \\
  \midrule
    \end{tabularx}}
\hspace*{0pt}%
\begin{minipage}{\textwidth}
\small \textcolor{myred_2}{* The best result for each dataset and metric is marked in \textbf{bold} and the second-best result is \underline{underlined}.}
\end{minipage}
\end{table*}

\subsection*{Evaluation strategies and metrics}
We conduct experiments across diverse datasets to evaluate MapDiff against state-of-the-art protein sequence design methods. We first perform evaluation on two popular benchmark datasets, CATH 4.2 and CATH 4.3 \cite{orengo1997cath}, using the same topology-based data split employed in previous works \cite{ingraham2019generative, hsu2022learning, gao2023pifold}. Besides the full test sets, we also study two subcategories of generated proteins: short proteins up to 100 residues in length and single-chain proteins (labelled with one chain in CATH). We use another two distinct datasets, TS50 \cite{li2014direct} and PDB2022 \cite{zhou2023prorefiner} to evaluate the zero-shot generalization of models. Furthermore, we study the foldability of the generated protein sequences via AlphaFold2 \cite{jumper2021highly} by comparing the discrepancy between the AlphaFold2-refolded structures and ground-truth native structures. \textcolor{red}{This is an \textit{in silico} evaluation rather than definitive proof that the designed sequences can fold into their intended structures.} The Experimental setting section provides detailed information and statistics about these datasets.

We evaluate the accuracy of generated sequences using three metrics: perplexity, recovery rate and native sequence similarity recovery (NSSR) \cite{loffler2017rosetta}. Perplexity measures the alignment between a model's predicted AA probabilities and the native AA types at each residue position. The recovery rate indicates the proportion of accurately predicted AAs in the protein sequence. The NSSR evaluates the similarity between the predicted and native residues via the Blocks Substitution Matrix (BLOSUM) \cite{henikoff1992amino}, where each residue pair contributes to a positive prediction if their BLOSUM score is greater than zero. We use BLOSUM42, BLOSUM62, BLOSUM80 and BLOSUM90 to account for AA similarities at four different cut-off levels for NSSR computation. To evaluate the foldability, i.e., the quality of refolded protein structures, we use six metrics: predicted local distance difference test (pLDDT), predicted aligned error (PAE), predicted template modelling (pTM), template modelling score (TM-score),  root mean square deviation (RMSD) and global distance test-total score (GDT-TS), \textcolor{myblue}{where pLDDT, PAE and pTM measure the confidence and reliability of predicted structures produced by AlphaFold2}, and TM-score, RMSD and GDT-TS measure the discrepancies between the predicted 3D structures and their native counterparts. \textcolor{red}{Supplementary S9 provides the technical details for these metrics.}

\subsection*{Sequence recovery performance}

Firstly, \textcolor{myblue}{we evaluate MapDiff's sequence recovery with uniform or marginal priors against state-of-the-art baselines on the CATH datasets.} Table \ref{tab:results_cath} presents the prediction perplexity and median recovery rate on the full test set, along with short and single-chain subsets. The results demonstrate that MapDiff achieves the best performance across different metrics and subsets of data, highlighting its effectiveness in generating valid protein sequences. Specifically, we can observe that: (1) \textcolor{myblue}{MapDiff achieves a recovery rate of 61.03\% and 60.86\% on the full CATH 4.2 and CATH 4.3 test sets, substantially outperforming the baselines by 7.74\% and 7.20\%, respectively.} Furthermore, \textcolor{myblue}{MapDiff shows recovery improvements of 8.20\% and 6.61\% on the short and single-chain test sets of CATH 4.2.} (2) MapDiff consistently achieves the lowest perplexity compared to previous methods and produce high-confidence probability distribution to facilitate accurate predictions. (3) MapDiff is a highly accurate IPF model that operates independently of external knowledge. In some of the compared baselines, external knowledge sources, such as additional training data or protein language models, are utilized to enhance prediction accuracy. Due to its well-designed architecture and diffusion-based generation mechanism, MapDiff effectively utilizes limited training data to capture relevant patterns to achieve superior generalizability. \textcolor{red}{(4) MapDiff's performance is largely unaffected by the choice of prior distribution. Therefore, we use the marginal prior \cite{vignac2022digress} in our experiments, as it is data-driven and better aligns with the true amino acid distribution.}

\begin{figure*}[t!]
    \begin{center}
    \includegraphics[width=1\textwidth]{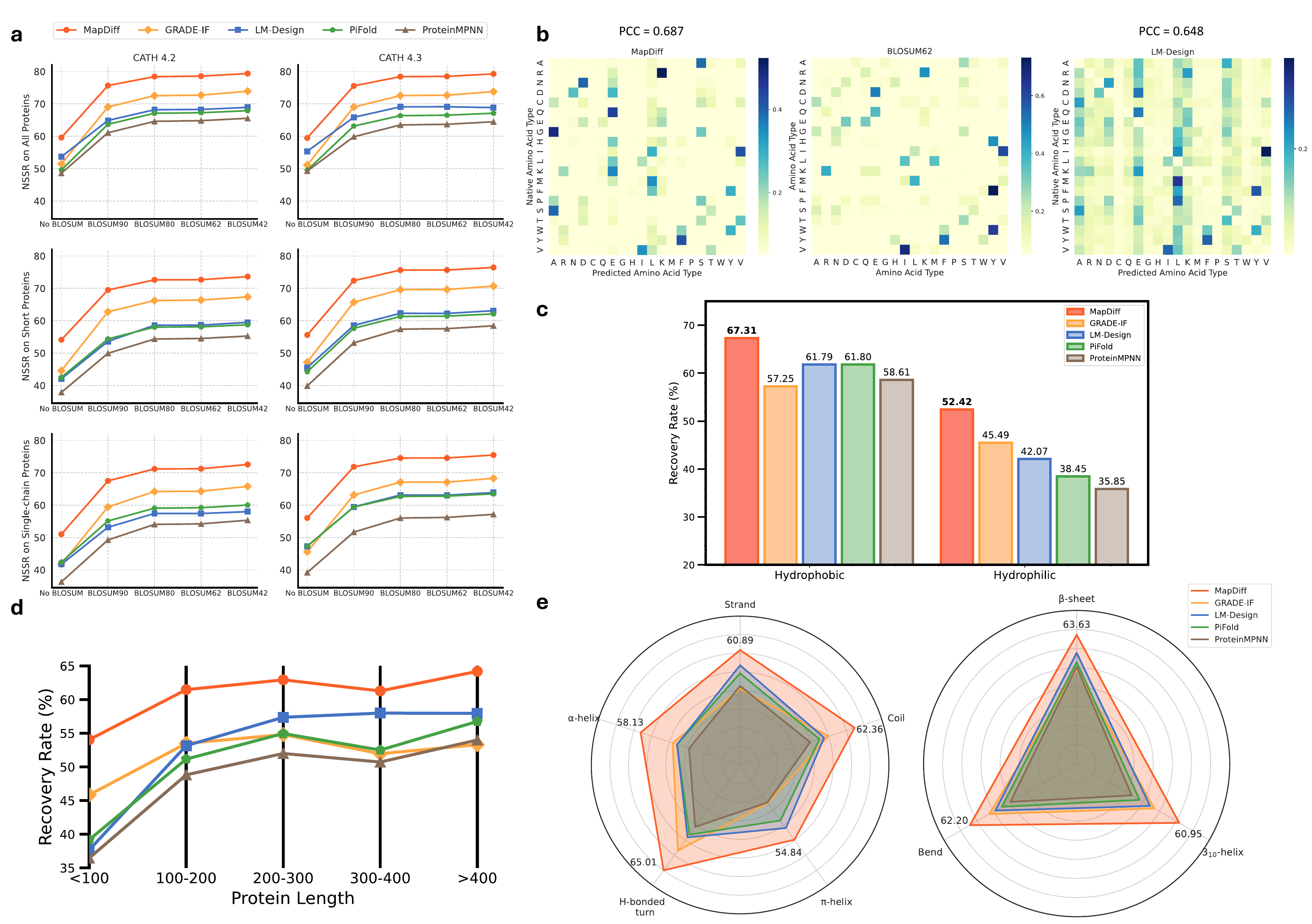}
    \end{center}
    \caption{\textbf{Model performance comparison and sensitivity analysis across different scenarios on the CATH datasets.} \textbf{(a)} NSSR scores for MapDiff and baseline methods \cite{yi2024graph, gao2023pifold, lmdesign, dauparas2022robust} on the full test sets and the short and single-chain protein subsets for four different BLOSUM matrices and no BLOSUM matrix. \textbf{(b)} \textcolor{myblue}{Sum-normalized confusion matrices for MapDiff (left) and LM-Design (right) predictions, and the softmax-normalized native BLOSUM62 matrix (middle). Darker colors indicate a higher predicted likelihood, with diagonal elements masked. The Pearson correlation coefficient (PCC) between the predicted matrix and BLOSUM62 quantifies their similarity and linear correlation.} \textbf{(c)} Breakdown of the recovery rates into hydrophilic and hydrophobic residues. \textbf{(d)} Median sequence recovery rates across different protein lengths. \textbf{(e)} Residue recovery performance across different secondary structures visualized in two groups for clarity, \textcolor{myblue}{defined using the DSSP algorithm \cite{kabsch1983dictionary}. Coils correspond to regions without regular secondary structures and are considered disordered regions. Bends and H-bonded turns are regarded as less ordered regions due to their flexibility and transient nature. For these regions, MapDiff outperforms the baselines substantially.}}
    \label{exp_fig_1}
    \vspace{-.5em}
\end{figure*}

\textcolor{myblue}{We further study model performance across different scenarios.} Fig. \ref{exp_fig_1}a presents the mean NSSR scores for MapDiff and the baselines on the CATH datasets. MapDiff consistently achieves the best NSSR scores across different test sets. \textcolor{myblue}{Fig. \ref{exp_fig_1}b compares the confusion matrices of MapDiff and LM-Design with the native BLOSUM62 matrix on CATH 4.2. For clearer visualization and comparison, normalized these matrices to the [0,1] probability range, with the diagonal elements masked. The confusion matrix denotes proportions for specific combinations of actual and predicted amino acid types, with darker cells indicating greater proportion. Many non-diagonal darker cells in MapDiff highlight the alignment between closely related residue pairs, as defined by the BLOSUM62 matrix, indicating that MapDiff can effectively capture the homologous substitutions between residues. Additionally, MapDiff's higher correlation with BLOSUM62 than LM-Design suggests a stronger alignment with substitution preferences.}

Figures \ref{exp_fig_1}c and \ref{exp_fig_1}e show the sequence recovery performance across different amino acid types, as well as eight secondary structures. Notably,  MapDiff is the only model achieving over 50\% recovery rate in predicting hydrophobic amino acids and substantial improvements in recovering $\alpha$-helix and $\beta$-sheet secondary structures. Fig. \ref{exp_fig_1}d presents a sensitivity analysis of the recovery performance for varying protein lengths. For short proteins (less than 100 amino acids in length), several baselines show a marked decrease in performance. For instance, the recovery rate of LM-Design falls below 40\% for the short proteins. This could be due to the protein language model used in LM-Design being sensitive to protein length. \textcolor{myblue}{In contrast, MapDiff, which employs a mask-prior-guided denoising network and an iterative denoising process,} consistently outperforms all baselines and maintains high performance across all protein lengths.

To validate the zero-shot transferability of our method, we compare the model's performance on two independent test datasets, TS50 and PDB2022, which do not overlap with the CATH data, as shown in Table \ref{tab:results_TS50TS500}. The results demonstrate that MapDiff achieves the highest recovery and NSSR scores on both datasets. We can conclude that, even though LM-Design reaches a high recovery (66\%) that is approaching our method on PDB2022, the performance gap widens on NSSR62 and NSSR90. In contrast, GRADE-IF and MapDiff can generalize better while considering the possibility of similar residue substitution. This suggests that diffusion-based models more effectively capture residue similarity in IPF prediction. \textcolor{myblue}{For the TS50 dataset, MapDiff substantially improves state-of-the-art methods by 6.33\% on NSSR62, and is the best model, achieving a recovery rate of 68\%.}

\begin{table}[t]
   \centering
   \setlength{\tabcolsep}{2pt}
   \caption{\textbf{Transferability}: zero-shot performance comparison on transferability from CATH to PDB2022 and TS50 datasets. We report the test results of models when trained on CATH 4.2 and CATH 4.3, with the results for CATH 4.3 in parentheses. \textbf{Foldability}: foldability comparison for the generated sequences on the CATH 4.2 test set using AlphaFold2. The results are presented as mean±standard deviation.}
   \label{tab:results_TS50TS500}
   \resizebox{1\linewidth}{!}{%
   \begin{tabular}{p{1cm}lp{2.5cm}p{2.5cm}p{2.5cm}p{2.5cm}p{2.5cm}p{2.5cm}}
   \toprule
   & \multirow{2}{*}{\bf Models} & 
    \multicolumn{3}{c}{\bf PDB2022 }  &
    \multicolumn{3}{c}{\bf TS50} \\
   \cmidrule[0.3pt](lr){3-5} \cmidrule[0.3pt](lr){6-8}
   &  & \bf Recovery (\textcolor{myblue}{$\uparrow$})  & \bf NSSR62 (\textcolor{myblue}{$\uparrow$})  & \bf NSSR90 (\textcolor{myblue}{$\uparrow$})  & \bf Recovery (\textcolor{myblue}{$\uparrow$})  & \bf NSSR62 (\textcolor{myblue}{$\uparrow$})  & \bf NSSR90 (\textcolor{myblue}{$\uparrow$})\\
   \midrule
   \multirow{5}{*}{\rotatebox[origin=c]{90}{\small \bf Transferability}} & ProteinMPNN \cite{dauparas2022robust}   & 56.75 (56.65) & 72.50 (72.59)  & 69.96 (69.95)  & 52.34 (51.80) & 70.31 (70.13) &  66.77 (66.80)  \\ 
    & PiFold \cite{gao2023pifold}   & 60.63 (60.26) & 75.55 (75.30) &  72.96 (72.86)   & \underline{58.39} (58.90)   & 73.55 (74.52) & 70.33 (71.33)   \\ 
    & LM-Design \cite{lmdesign} & \underline{66.03} (\underline{66.20}) & \underline{79.55} (\underline{80.12}) & \underline{77.60} (\underline{78.20}) & 57.62 (58.27) & 73.74 (75.69)  &  71.22 (73.12)  \\
   & GRADE-IF \cite{yi2024graph}  & 58.09 (58.35) & 77.44 (77.51) & 74.57 (74.97)  &  57.74 (\underline{59.27})  & \underline{77.77} (\underline{79.11})  &  \underline{74.36} (\underline{76.24}) \\ 
   & MapDiff  & \textbf{68.03} (\textbf{68.00}) & \textbf{84.19} (\textbf{84.30}) & \textbf{82.13} (\textbf{82.29})  &  \textbf{68.76} (\textbf{69.77}) & \textbf{84.10} (\textbf{85.27})   & \textbf{81.76} (\textbf{83.08}) \\
   \midrule
   & \multirow{2}{*}{\bf Models} & 
    \multicolumn{6}{c}{\bf CATH 4.2} \\
   \cmidrule[0.3pt](lr){3-5} \cmidrule[0.3pt](lr){5-8}
   &  & \bf pLDDT ($\uparrow$) & \bf PAE ($\downarrow$) & \bf PTM ($\uparrow$) & \bf TM-score ($\uparrow$) & \bf RMSD ($\downarrow$) & \bf GDT-TS ($\uparrow$) \\ \midrule
    \multirow{5}{*}{\rotatebox[origin=c]{90}{\small \bf Foldability}} & ProteinMPNN \cite{dauparas2022robust}    & 87.13$\pm$9.79        & 5.85$\pm$3.17   & 77.42$\pm$14.96    & 86.27$\pm$16.32            & \underline{3.08$\pm$4.25} & \underline{85.08$\pm$15.53}  \\
     & PiFold \cite{gao2023pifold}        & 87.42$\pm$9.82       & 5.81$\pm$3.22 &77.75$\pm$15.03       & \underline{86.56$\pm$16.21}            & 3.10$\pm$4.29  & 85.47$\pm$15.49 \\
    & LM-Design \cite{lmdesign}      & \underline{88.04$\pm$9.00}        & \underline{5.78$\pm$3.27} & \underline{78.00$\pm$14.80}       & 85.36$\pm$16.98            & 3.54$\pm$5.00  & 84.08$\pm$16.45 \\
    & GRADE-IF \cite{yi2024graph} & 85.32$\pm$9.27        & 6.30$\pm$3.10 & 75.63$\pm$13.80     & 85.80$\pm$14.93        & 3.11$\pm$3.96  & 83.37$\pm$14.43 \\
    & MapDiff & \textbf{88.63$\pm$8.27}        & \textbf{5.42$\pm$2.76} & \textbf{79.00$\pm$13.04}       & \textbf{88.77$\pm$13.48} & \textbf{2.57$\pm$3.50} & \textbf{87.75$\pm$13.24} \\
   \bottomrule
   \end{tabular}
   }
\hspace*{0pt}%
\begin{minipage}{\textwidth}
\small \textcolor{myred_2}{* The best result for each dataset and metric is marked in \textbf{bold} and the second-best result is \underline{underlined}.}
\end{minipage}
\end{table}

\begin{figure*}[!t]
    \begin{center}
    \includegraphics[width=\textwidth]{figures/fold_visual_v6.pdf}
    \end{center}
     \caption{\textbf{Comparison of three refolded structures (left) and the respective model-designed sequences (right) for proteins with PDB IDs 1NI8, 2HKY, and 2P0X.} \textbf{(a)} Refolded tertiary structure visualization of the sequences designed by three models MapDiff (red), GRADE-IF (orange) and LM-Design (blue). The refolded structures are generated by AlphaFold2 and superposed against the ground-truth structures (purple). For each model and structure, the recovery rate and RMSD value are indicated for foldability comparison. \textbf{(b)} The alignment of the three native sequences and the respective model-designed sequences. The results are shown with secondary structure elements marked below each sequence: \textcolor{myblue}{$\alpha$-helices are shown in red cylinders, $\beta$-strands in blue arrows, and loops and disordered regions are unmarked. For the native proteins, the secondary structures were derived from their source PDB files. For the predicted proteins, the secondary structures were assigned by first identifying all inter-backbone hydrogen bonds and then searching for hydrogen-bonding patterns that represent helices and strands.} The refolded structures and sequence alignments are visualized using the Schrödinger Maestro software \cite{schrodinger2023maestro}.
     \textcolor{red}{\textbf{(c)} Recovery rates for loops and disordered regions (left panel) and $\alpha$-helix and $\beta$-strand regions (right panel) across three structures. Bars indicate the recovery rates of three methods (MapDiff, Grade-IF, and LM-Design). The percentage composition of regions for each structure is provided below the panel titles. MapDiff consistently achieves the highest recovery rates across different categories of regions for the three structures, with an average improvement of 5.1\% in loops and disordered regions and 13.4\% in $\alpha$-helix and $\beta$-strand regions compared to Grade-IF.
     \textbf{(d)} Jaccard region intersections between the predicted and ground-truth structures for loops and disordered regions (left panel) versus $\alpha$-helix and $\beta$-strand regions (right panel). The Jaccard index measures the fraction of the overlap between two sets, and the results demonstrate that MapDiff achieves the highest score across both categories of regions.}
     }
    \label{fold_visual}
    \vspace{5mm}
\end{figure*}

\subsection*{Foldability of generated protein sequences}
Foldability is a crucial property that evaluates whether a protein sequence can fold into the desired structure. In this study, we evaluate the foldability of generated protein sequences by predicting their structures with AlphaFold2 and comparing the discrepancies against the native crystal structures. Table \ref{tab:results_TS50TS500} presents six foldability metrics for the 1,120 structures in the CATH 4.2 test set. The results indicate that the generated protein sequences by MapDiff exhibit superior foldability, the highest confidence, and minimal discrepancy compared to their native structures. Notably, the foldability and sequence recovery results do not always positively correlate. For instance, while ProteinMPNN performs poorly in sequence recovery, it achieves the best RMSD among baseline methods. Therefore, it is essential to comprehensively evaluate IPF models from both sequence and structure perspectives. \textcolor{red}{Supplementary S2 and Supplementary Fig. 2 present analysis of the right-skewed RMSD distribution \cite{log_normal}.}

In Fig. \ref{fold_visual}a, we illustrate exemplary 3D structures refolded by AlphaFold2 from IPF-derived sequences generated by MapDiff, GRADE-IF, and LM-Design for three different protein folds (PDB ID: 1NI8 \cite{bloch2003h}, 2HKY \cite{huang2007flexible}, 2P0X \cite{mansy2007structure}) with a pre-selected monomer pTM prediction argument. In addition to estimating the sequence recovery rate and foldability of derived 3D structures using the RMSD metric, we also inspect the alignment of native and generated sequences, including the agreement between refolded secondary structures and individual pairs of amino acids in Fig. \ref{fold_visual}b. \textcolor{red}{Figs. \ref{fold_visual}c and \ref{fold_visual}d present quantitative analyses of performance on different regions.}

The first example is a 46 amino acid-long monomer of the 1NI8 structure (purple) representing an N-terminal fragment of the H-NS dimerization domain, a protein composed of three $\alpha$ helices that is involved in structuring the chromosome of Gram-negative bacteria, hence acting as a global regulator for the expression of different genes \cite{bloch2003h}. Two monomers form a homodimer which requires the presence of \myrev{$K5$, $R11$, $R{14}$, $R{18}$, and $K{31}$} residues to engage in the prokaryotic DNA binding. 
MapDiff (red) managed to retrieve two out of the three $\alpha$ helices, with an interhelical turn present at the same position as in the original structure (A17-R18), whereas GRADE-IF (orange) and LM-Design (blue) models only consisted of a single continuous $\alpha$ helix. Moreover, MapDiff and GRADE-IF obtained four out of five (\myrev{$K5$, $R{11}$, $R{14}$, and $R{18}$}) amino acids required for DNA binding and LM-Design obtained none. MapDiff and LM-Design generate glutamic acid (E) and GRADE-IF isoleucine (I), which, in comparison to the corresponding positively charged \myrev{$K{31}$} in the original structure, are negatively charged and neutral residues, respectively. The single continuous $\alpha$ helix displayed by GRADE-IF and LM-Design AlphaFold2 models hence produces much worse RMSD values (14.5 \AA{} and 14.2 \AA, respectively) than the MapDiff model, which retrieved two helices at the right positions (RMSD = 4.6 \AA). Consistent with this, MapDiff obtained a 10\% higher recovery rate than GRADE-IF and LM-Design.

The second example is the 2HKY structure of 109 amino-acid long human ubiquitous ribonuclease 7 (hRNase7), rich in positively charged residues, that possesses antimicrobial activity \cite{huang2007flexible}. This $\alpha$/$\beta$ mixed protein contains 22 cationic residues (18 K and 4 R) distributed into three surface-exposed clusters which promote binding to the bacterial membrane, thus rendering it permeable, which consequently elicits membrane disruption and death. In addition, it contains four disulfide bridges (
\myrev{$C{24}-C{82}$, $C{38}-C{92}$, $C{56}-C{107}$, and $C{63}-C{70}$}
) critical for its secondary and tertiary structure, three of which were successfully retrieved by MapDiff, whereas no cysteines were found in either GRADE-IF or LM-Design sequences. Furthermore, all secondary structure elements were nearly entirely recovered by MapDiff, unlike GRADE-IF and LM-Design solutions which contained little resemblance to the native structure, particularly in the C-terminus half. These structural findings were reflected in a fair recovery rate of 40.3\% and a RMSD value of 5.0 Å for MapDiff, which was considerably better than in GRADE-IF and LM-Design structures (14.0 \AA{} and 12.6 \AA, respectively).

A third example displays AlphaFold2 refolded structures obtained from generated sequences with relatively low recovery rates that used 2P0X structure of an optimized non-biological (de novo) ATP-binding protein as a template \cite{mansy2007structure}. Here, MapDiff retrieved all detected secondary structure elements, except for the C-terminus $\beta$-strand which was replaced by a loop. LM-Design was the second best with an $\alpha$ helix substituting the aforementioned $\beta$-strand. Even if nearly all secondary structure elements were retrieved by both MapDiff and LM-Design AlphaFold2 models, MapDiff model obtained by far the best RMSD (3.3 \AA{} as opposed to 8.8 \AA). Despite having a better recovery rate than LM-Design, GRADE-IF generated a sequence that folded poorly compared to the experimentally confirmed structure (15.0 \AA).

In these cases, MapDiff achieved low RMSD values to successfully replicate the majority of secondary structure elements elucidated through experiments, including other structural features such as the disulfide bonds (2HKY) or positively charged residues that were suspected to participate in protein function (1NI8). In contrast, Grade-IF and LM-Design predicted sequences that not only had lower recovery rates than MapDiff but also exhibited partially or entirely absent secondary structure elements, as shown by the experimentally derived 3D structures, resulting in substantially worse RMSDs. Although the structures predicted by AlphaFold2 cannot entirely substitute the structural elucidation by experimental techniques such as X-ray or NMR (nuclear magnetic resonance), they provide the first glance at the foldability potential of de novo generated protein sequences by IPF models. A natural next step in future work would be to express the de novo designed protein sequences and experimentally determine their tertiary structures. 

\textcolor{red}{Supplementary S1 and Supplementary Fig. 1 study the closest training structures and sequences of the three examples. The highest TM-scores for 1NI8, 2HKY, and 2P0X from structures in the training set are 0.57 (1A7W), 0.25 (1V88) and 0.33 (1WIM), respectively, indicating that there are no highly similar structures during training. Similarly, the highest BLAST \cite{altschul1990basic} bit-scores for sequences in the training set are 23.1 (4ZEO), 26.6 (2BM8) and 24.6 (3MSR), respectively, indicating that no highly similar sequences are present during training.}

\subsection*{Model analysis and ablation study}
We perform analysis and ablation study to assess the effectiveness of key components in MapDiff. We focus on investigating the contributions of edge feature updating, node coordinate updating and global context learning within the base sequence predictor (G-EGNN) to the model performance. Additionally, we examine the impact of the mask ratio adapter and the pre-trained IPA network in the residue refinement module on the predictions. As shown in Table \ref{tab:ablation}, we study five variants of MapDiff, each with different key components removed, \textcolor{myblue}{and compare their results on the CATH 4.2 test set. The results show that each component positively enhances the sequence recovery and foldability performance.} \textcolor{myblue}{For instance, the IPA-based refinement mechanism (variant 5) achieves the most substantial sequence improvement, increasing recovery by 4.47\%, while the global context learning and coordinate updating (variants 2 and 4) in G-EGNN improve the recovery by 1.17\% and 0.77\%, respectively.} \textcolor{red}{The impact on foldability increases with sequence recovery performance but remains less pronounced, indicating that AlphaFold2 is robust to these variations and predicts stable protein folds. In addition, Supplementary S7 and Supplementary Fig. 3 analyze MapDiff's sensitivity to the number of Monte-Carlo samples and DDIM skipping steps.}

\begin{table}[t]
   \small
   \centering
   \caption{Ablation study of the denoising network modules in MapDiff. We study five model variants and investigate how much sequence recovery and foldability metrics decrease when key components are removed on CATH 4.2. The best result for each metric is highlighted in bold.}
    \begin{tabular}{llllllll}
    \toprule
    Module                                                                             & Component          & \multicolumn{1}{c}{MapDiff} & \multicolumn{1}{c}{Variant 1} & \multicolumn{1}{c}{Variant 2} & \multicolumn{1}{c}{Variant 3} & \multicolumn{1}{c}{Variant 4} & \multicolumn{1}{c}{Variant 5} \\ \hline
    \multirow{3}{*}{G-EGNN} & EdgeUpdate        & \multicolumn{1}{c}{\cmark}       & \multicolumn{1}{c}{\cmark}         & \multicolumn{1}{c}{\cmark}         &                               & \multicolumn{1}{c}{\cmark}         & \multicolumn{1}{c}{\cmark}      \\
                                                                                       & CoordinateUpdate  & \multicolumn{1}{c}{\cmark}       & \multicolumn{1}{c}{\cmark}         & \multicolumn{1}{c}{\cmark}         &                               &          &   \multicolumn{1}{c}{\cmark}      \\
                                                                                       & GlobalContext     & \multicolumn{1}{c}{\cmark}       & \multicolumn{1}{c}{\cmark}         &                               & \multicolumn{1}{c}{\cmark}         & \multicolumn{1}{c}{\cmark}         & \multicolumn{1}{c}{\cmark}        \\ \hline
    \multirow{2}{*}{Refinement}    & MaskAdapter & \multicolumn{1}{c}{\cmark}       &                               & \multicolumn{1}{c}{\cmark}         & \multicolumn{1}{c}{\cmark}         &    \multicolumn{1}{c}{\cmark}                           &               \\
                                                                                      & IPA network                & \multicolumn{1}{c}{\cmark}       & \multicolumn{1}{c}{\cmark}         & \multicolumn{1}{c}{\cmark}         & \multicolumn{1}{c}{\cmark}         & \multicolumn{1}{c}{\cmark}         &      \\ \hline
    \multirow{3}{*}{Sequence}                                                           & Recovery ($\uparrow$, \%)             &    \multicolumn{1}{c}{\textbf{60.93}}                         &     \multicolumn{1}{c}{58.64}                           &  \multicolumn{1}{c}{59.76}                             &       \multicolumn{1}{c}{58.38}                        &         \multicolumn{1}{c}{60.16}                       &          \multicolumn{1}{c}{56.46}                                     \\
                                                                                      & NSSR62 ($\uparrow$, \%)            &   \multicolumn{1}{c}{\textbf{78.57}}                          &   \multicolumn{1}{c}{76.73}                            &    \multicolumn{1}{c}{77.32}                           &     \multicolumn{1}{c}{77.04}                          &        \multicolumn{1}{c}{77.80}                       &           \multicolumn{1}{c}{75.69}                                 \\
                                                                                      & NSSR90 ($\uparrow$, \%)             &   \multicolumn{1}{c}{\textbf{75.66}}                          &    \multicolumn{1}{c}{73.52}                           &    \multicolumn{1}{c}{74.82}                           &       \multicolumn{1}{c}{74.24}                        &           \multicolumn{1}{c}{75.02}                    &           \multicolumn{1}{c}{72.58}                            \\ \hline
    \multirow{6}{*}{Foldability}
    & pLDDT ($\uparrow$)          &   \multicolumn{1}{c}{\textbf{88.63}}                           &   \multicolumn{1}{c}{88.08}                            &    \multicolumn{1}{c}{88.24}                           &     \multicolumn{1}{c}{87.95}                          &        \multicolumn{1}{c}{88.30}                       &           \multicolumn{1}{c}{86.95}                                 \\
                                                                                      & PTM ($\uparrow$)          &   \multicolumn{1}{c}{\textbf{79.00}}                          &   \multicolumn{1}{c}{78.42}                            &    \multicolumn{1}{c}{78.61}                           &     \multicolumn{1}{c}{78.24}                          &        \multicolumn{1}{c}{78.74}                       &           \multicolumn{1}{c}{77.20}                                 \\
                                                                                      & PAE ($\downarrow$)           &   \multicolumn{1}{c}{\textbf{5.42}}                          &   \multicolumn{1}{c}{5.57}                            &    \multicolumn{1}{c}{5.54}                           &     \multicolumn{1}{c}{5.62}                          &        \multicolumn{1}{c}{5.49}                       &           \multicolumn{1}{c}{5.86}                                 \\
                                                                                      & TM-Score ($\uparrow$, \%)           &   \multicolumn{1}{c}{\textbf{88.77}}                          &   \multicolumn{1}{c}{88.25}                            &    \multicolumn{1}{c}{88.47}                           &     \multicolumn{1}{c}{88.16}                          &        \multicolumn{1}{c}{88.58}                       &           \multicolumn{1}{c}{87.50}                                 \\
                                                                                      & GDT-TS ($\uparrow$, \%)           &   \multicolumn{1}{c}{\textbf{87.75}}                          &   \multicolumn{1}{c}{87.12}                            &    \multicolumn{1}{c}{87.40}                           &     \multicolumn{1}{c}{86.96}                          &        \multicolumn{1}{c}{87.53}                       &           \multicolumn{1}{c}{85.72}                                 \\ 
                                                                                      & RMSD ($\downarrow$)           &   \multicolumn{1}{c}{\textbf{2.57}}                          &   \multicolumn{1}{c}{2.67}                            &    \multicolumn{1}{c}{2.65}                           &     \multicolumn{1}{c}{2.65}                          &        \multicolumn{1}{c}{2.53}                       &           \multicolumn{1}{c}{2.76}                                 \\\hline                                                                                     
    \multirow{1}{*}{Summary}                                                          & Change                &      \multicolumn{1}{c}{-}                        &       \multicolumn{1}{c}{$\downarrow\downarrow$}                           &           \multicolumn{1}{c}{$\downarrow$}                       &      \multicolumn{1}{c}{$\downarrow\downarrow$}                            &       \multicolumn{1}{c}{$\downarrow$}                           &       \multicolumn{1}{c}{$\downarrow\downarrow\downarrow$}                  \\ \bottomrule
    \end{tabular}
    \label{tab:ablation}
\hspace*{0pt}%
\begin{minipage}{\textwidth}
\small \textcolor{myred_2}{*The best result for each metric is highlighted in bold.}
\end{minipage}
\end{table}

\section*{Discussion}
In this work, we present MapDiff, a mask-prior-guided denoising diffusion framework for structure-based protein design. Specifically, we regard IPF prediction as a discrete denoising diffusion problem, and develop a graph-based denoising network to capture structural information and residue interactions. At each denoising step, we utilize a G-EGNN module to generate clean sequences from input structures and a pre-trained IPA module to refine low-confidence residues, ensuring reliable denoising trajectories. Moreover, we integrate DDIM with Monte-Carlo dropout to accelerate generative sampling and enhance uncertainty estimation. Experiments demonstrate that MapDiff consistently outperforms the state-of-the-art IPF models across multiple benchmarks and scenarios. At the same time, the generated protein sequences exhibit a high degree of similarity to their native counterparts. Even in cases where the overall sequence similarity was low, these sequences can often refold into their native structures, as demonstrated by the AlphaFold2-refolded models. We also conduct a comprehensive ablation study to analyze the importance of different model components for the prediction results. MapDiff demonstrates transferability and robustness in generating novel protein sequences, even with limited training data. Promising future directions include: verifying the applicability of MapDiff in practical domains such as de novo antibody design and protein engineering, \textcolor{red}{incorporating predicted structures from structure prediction models as external data for incremental training, integrating physics-informed constraints, leveraging sequential evolutionary knowledge from protein language models to further refine residue predictions, and further validating the foldability of the designed sequences by conducting folding simulations or molecular dynamics simulations.}

\section*{Methods}
\subsection*{Discrete denoising diffusion models}
Denoising diffusion models are a class of deep generative models trained to create new samples by iteratively denoising sampled noise from a prior distribution. The training stage of a diffusion model consists of a forward diffusion process and a reverse denoising process. Given an original data distribution $q(\mathbf{x}_0)$, the forward diffusion process gradually corrupts a data point $\mathbf{x}_0 \sim q(\mathbf{x}_0)$ into a series of increasingly noisy data points $\mathbf{x}_{1:T} = \mathbf{x}_1, \mathbf{x}_2, \cdots, \mathbf{x}_T$ over $T$ time steps. This process follows a Markov chain, where $q(\mathbf{x}_{1:T} \mid \mathbf{x}_0) = \prod_{t=1}^{T} q(\mathbf{x}_t \mid \mathbf{x}_{t-1})$. Conversely, the reverse denoising process, denoted as $p_{\theta}(\mathbf{x}_{0:T}) = p(\mathbf{x}_T)\prod_{t=1}^{T}p_{\theta}(\mathbf{x}_{t-1} \mid \mathbf{x}_t)$, aims to progressively reduce noise towards the original data distribution $q(\mathbf{x}_0)$ by predicting $\mathbf{x}_{t-1}$ from $\mathbf{x}_{t}$. The initial noise $\mathbf{x}_{T}$ is sampled from a pre-defined prior distribution $p(\mathbf{x}_T)$, and the denoising inference $p_{\theta}$ can be parameterized by a learnable neural network. While the diffusion and denoising processes are agnostic to the data modality, the choice of prior distributions and Markov transition operators varies between continuous and discrete spaces.

In this work, we follow the settings of the discrete denoising diffusion proposed by Austin et al. (2021) \cite{austin2021structured} and Clement et al. (2023) \cite{vignac2022digress}. In contrast to typical Gaussian diffusion models that operate in continuous state space, discrete denoising diffusion models introduce noise to categorical data using transition probability matrices in discrete state space. Let $x_t \in \{1, \cdots, K\}$ denote the categorical data with $K$ categories and its one-hot encoding represented as $\mathbf{x}_t \in \mathbb{R}^{K}$. At time step $t$, the forward transition probabilities can be denoted by a matrix $\mathbf{Q}_t \in \mathbb{R}^{K \times K}$, where $[\mathbf{Q}_t]_{ij} = q(x_t = j \mid x_{t-1} = i)$ is the probability of transitioning from category $i$ to category $j$. Therefore, the discrete transition kernel in the diffusion process is defined as: 

\begin{align}
    & q(\mathbf{x}_t \mid \mathbf{x}_{t-1}) = \mathrm{Cat}(\mathbf{x}_t; \mathbf{p} = \mathbf{x}_{t-1}\mathbf{Q}_t),
    \label{eq_trans_1}\\
    & q(\mathbf{x}_t \mid \mathbf{x}_{0}) = \mathrm{Cat}(\mathbf{x}_t; \mathbf{p} = \mathbf{x}_{0}\overline{\mathbf{Q}}_t), \ \mathrm{with} \ \overline{\mathbf{Q}}_t = \mathbf{Q}_1\mathbf{Q}_2\cdots\mathbf{Q}_t,
    \label{eq_trans}
\end{align}

\noindent where $\mathrm{Cat}(\mathbf{x; \mathbf{p}})$ represents a categorical distribution over $\mathbf{x}_t$ with probabilities determined by $\mathbf{p} \in \mathbb{R}^{K}$. As the diffusion process has a Markov chain, the transition matrix from $\mathbf{x}_0$ to $\mathbf{\mathbf{x}_t}$ can be written as a closed form in equation (\ref{eq_trans}) with $\overline{\mathbf{Q}}_t = \mathbf{Q}_1\mathbf{Q}_2\cdots\mathbf{Q}_t$. This property enables efficient sampling of $\mathbf{x}_t$ at arbitrary time steps without recursively applying noise. Following the Bayesian theorem, the calculation of posterior distribution (with the derivation in Supplementary S3) from time step $t$ to $t-1$ can be written as:

\begin{equation}
q({\mathbf{x}_{t-1} \mid \mathbf{x}_{t}, \mathbf{x}_{0}}) \propto \mathbf{x}_t \mathbf{Q}_t^T \odot \mathbf{x}_0 \overline{\mathbf{Q}}_{t-1},
\label{post_dist}
\end{equation}

\noindent where $\odot$ is a Hadamard (element-wise) product. The posterior $q({\mathbf{x}_{t-1} \mid \mathbf{x}_{t}, \mathbf{x}_{0}})$  is equivalent to $q(\mathbf{x}_{t-1} \mid \mathbf{x}_{t})$ due its Markov property. Thus, the clean data $\mathbf{x}_0$ is introduced for denoising estimation and can be used as the target of the denoising neural network. \textcolor{myblue}{In MapDiff, we introduce two simple but effective choices for the transition matrix $\mathbf{Q}_t$: uniform transition \cite{austin2021structured} and marginal transition \cite{vignac2022digress}. The uniform transition is parametrized by $\mathbf{Q}_t = (1 - \beta_t)\mathbf{I} + \beta_t\mathbf{1}_K\mathbf{1}_K^T / K$, where $K = 20$ represents the number of native amino acid types and the noise schedule $\beta_t \in [0, 1]$. Similarly, the marginal transition is parameterized by $\mathbf{Q}_t = (1 - \beta_t) \mathbf{I} + \beta_t \mathbf{1}_K \mathbf{p}^T$, where $\mathbf{p} \in \mathbb{R}^{20}$ denotes the marginal probability distribution of AA types in the training data. All matrix values are strictly positive, and each row sums to one, ensuring the conservation of probability mass. Given these properties, along with the condition $\mathrm{lim}_{t \rightarrow T} \beta_t=1$, $q(\mathbf{x}_t)$ can converge to a stationary uniform or marginal distribution, regardless of the initial $\mathbf{x}_{0}$.}

\subsection*{Residue graph construction}

IPF prediction aims to generate a feasible AA sequence that can fold into a desired backbone structure. Given a target protein of length $L$, we present it as a proximity residue graph $\mathcal{G} = (\mathbf{X}, \mathbf{A}, \mathbf{E})$, where each node denotes an AA residue within the protein. The node features $\mathbf{X} = [\mathbf{X}^\mathrm{aa}, \mathbf{X}^\mathrm{pos}, \mathbf{X}^\mathrm{prop}]$ encodes the AA residue types, 3D spatial coordinates, and geometric properties. The adjacency matrix $\mathbf{A} \in \{0, 1\}^{N \times N}$ is constructed using the $k$-nearest neighbor algorithm. Specifically, each node is connected to a maximum of $k$ other nodes within a cutoff distance smaller than 30 Å. The edge feature matrix $\mathbf{E} \in \mathbb{R}^{M \times 93}$ illustrates the spatial and sequential relationships between the connected nodes. More details on the graph feature construction are provided in Supplementary S4. For sequence generation, we define a discrete denoising process on the types of noisy AA residues $\mathbf{X}^\mathrm{aa}_{t} \in \mathbb{R}^{N \times 20}$ at time $t$. Conditioned on the noise graph $\mathcal{G}_t$, this process is subject to iteratively refine noise $\mathbf{X}^\mathrm{aa}_{t}$ towards a clean $\mathbf{X}^\mathrm{aa}_{0} = \mathbf{X}^\mathrm{aa}$, which is predicted by our mask-prior-guided denoising network.

\subsection*{IPF denoising diffusion process}

\textbf{Discrete diffusion process.} In the diffusion process, we incrementally introduce discrete noise to the clean AA residues over a number of time steps $t \in \{{1, \cdots, T\}}$, \textcolor{myblue}{resulting in transforming the original data distribution to a simple uniform or marginal distribution.} Given a clean AA sequence $\mathbf{X}_0^\mathrm{aa} = \{\mathbf{x}_0^i \in \mathbb{R}^{1 \times 20} \mid 1 \leq i \leq N\}$, we utilize a cumulative transition matrix $\overline{\mathbf{Q}}_t$ to independently add noise to each AA residue at arbitrary step $t$:

\begin{align}
    & q(\mathbf{x}_t^i \mid \mathbf{x}_0^i) = \mathrm{Cat}(\mathbf{x}_t^{i}; \mathbf{p} = \mathbf{x}^{i}_{0}\overline{\mathbf{Q}}_t),  \mathrm{with} \ \overline{\mathbf{Q}}_t = \mathbf{Q}_1\mathbf{Q}_2\cdots\mathbf{Q}_t, \\
    & q(\mathbf{X}_t^\mathrm{aa} \mid \mathbf{X}_0^\mathrm{aa}) = \prod_{1 \leq i \leq N} q(\mathbf{x}_t^i \mid \mathbf{x}_0^i), 
\end{align}

\noindent where $\mathbf{Q}_t = (1 - \beta_t)\mathbf{I} + \beta_t\mathbf{1}_K\mathbf{1}_K^T / K$ as above, and $K$ denotes the number of native AA types (i.e. $K=20$). The weight of the noise, $\beta_t \in [0, 1]$ is determined by a common cosine schedule \cite{nichol2021improved}.

\noindent\textbf{Training objective of denoising network.} The denoising neural network, denoted as $\phi_{\theta}$,  is an essential component to reverse the noise process in diffusion models. In our framework, the network takes a noise residue graph  $\mathcal{G}_t = (\mathbf{X}_t, \mathbf{A}, \mathbf{E})$ as input and aims to predict the clean AA residues $\mathbf{X}^\mathrm{aa}_{0}$. Specifically, we design a mask-prior-guided denoising network $\phi_{\theta}$ to effectively capture inherent structural information and learn the underlying data distribution. To train the learnable network $\phi_{\theta}$, the objective is to minimize the cross-entropy loss between the predicted AA probabilities and the real AA types over all nodes.

\noindent\textbf{Reverse denoising process.} Once the denoising network has been trained, it can be utilized to generate new AA sequences through an iterative denoising process. In this study, we first use the denoising network $\phi_{\theta}$ to estimate the generative distribution $\hat{p}_{\theta}(\hat{\mathbf{x}}_0^i | \mathbf{x}_t^i)$ for each AA residue. Then the reverse denoising distribution $p_{\theta}(\mathbf{x}_{t-1}^i | \mathbf{x}_t^i)$ is parameterized by combining the posterior distribution with the marginalized network predictions as follows:

\begin{align}
    & p_{\theta}(\mathbf{x}_{t-1}^{i} | \mathbf{x}_t^{i}) \propto \sum_{\hat{\mathbf{x}}_0^{i}} q({\mathbf{x}_{t-1}^{i} \mid \mathbf{x}_{t}^{i}, \hat{\mathbf{x}}_{0}^{i}}) \hat{p}_{\theta}(\hat{\mathbf{x}}_0^{i} \mid \mathbf{x}_t^{i}), \\
    & p_{\theta}(\mathbf{X}_{t-1}^\mathrm{aa} \mid \mathbf{X}_t^\mathrm{aa}) = \prod_{1 \leq i \leq N} p_{\theta}(\mathbf{x}_{t-1}^i \mid \mathbf{x}_{t}^i), 
\end{align}

\noindent where $\hat{\mathbf{x}}_0^i$ represents the predicted probability distribution for the $i$-th residue $\mathbf{x}_0^i$. The posterior distribution is defined as:

\begin{align}
    q({\mathbf{x}_{t-1}^{i} \mid \mathbf{x}_{t}^{i}, \hat{\mathbf{x}}_{0}^{i}}) &= \frac{q(\mathbf{x}_t^{i} \mid \mathbf{x}_{t-1}^{i}, \hat{\mathbf{x}}_{0}^{i}) q(\mathbf{x}_{t-1}^{i} \mid \hat{\mathbf{x}}_{0}^{i})}{q(\mathbf{x}_{t}^{i} \mid \hat{\mathbf{x}}_{0}^{i})}, \nonumber \\
    &= \mathrm{Cat}\left(\mathbf{x}_{t-1}^{i}; \mathbf{p} = \frac{\textbf{x}_t^i \textbf{Q}_t^T \odot \hat{\mathbf{x}}_0^i \overline{\mathbf{Q}}_{t-1}}{\hat{\mathbf{x}}^{i}_0 \overline{\mathbf{Q}}_t (\mathbf{x}_t^i)^T}\right).
    \label{post_res}
\end{align}

By applying the reverse denoising process, the generation of less noisy $\mathbf{X}_{t-1}^\mathrm{aa}$ from $\mathbf{X}_{t}^\mathrm{aa}$ is feasible (derivation in Supplementary S3). The denoised result is determined by the predicted residues from the denoising neural network, as well as the predefined transition matrices at steps $t$ and $t-1$. To generate a new AA sequence, the complete generative process begins with a random noise from the independent prior distribution $p(\mathbf{x}_T)$. The initial noise is then iteratively denoised at each time step using the reverse denoising process, gradually converging to a desired sequence conditioned on the given graph $\mathcal{G}$. 

\noindent \textbf{DDIM with Monte-Carlo dropout.}
Although discrete diffusion models have demonstrated impressive generation ability in many fields, the generative process suffers from two limitations that hinder their success in IPF prediction. Firstly, the generative process is inherently computationally inefficient due to the numerous denoising steps involved, which require a sequential Markovian forward pass for the iterative generation. Secondly, the categorical distribution utilized for denoising sampling lacks sufficient uncertainty estimation. Many studies indicate that the logits produced by deep neural networks do not accurately represent the true probabilities. Typically, the predictions tend to be overconfident, leading to a discrepancy between the predicted probabilities and actual distribution. As the generative process iteratively draws samples from the estimated categorical distribution, insufficient uncertainty estimation will accumulate sampling errors and result in unsatisfactory performance.

To accelerate the generative process and improve uncertainty estimation, we propose a novel discrete sampling method by combining DDIM with Monte-Carlo dropout. DDIM, known as Denoising Diffusion Implicit Model \cite{song2021denoising}, is a widely used method that improves the generation efficiency of diffusion models in continuous space. It defines the generative process as the reverse of a deterministic and non-Markovian diffusion process, making it possible to skip certain denoising steps during generation. As discrete diffusion models possess analogous properties, Yi et al. (2023) \cite{yi2024graph} extended DDIM into discrete space for IPF prediction. Similarly, we define the discrete DDIM sampling to the posterior distribution as follows:

\begin{align}
    q({\mathbf{x}_{t-k}^{i} \mid \mathbf{x}_{t}^{i}, \hat{\mathbf{x}}_{0}^{i}}) 
    = \mathrm{Cat}\left(\mathbf{x}_{t-k}^{i}; \mathbf{p} = \frac{\textbf{x}_t^i \textbf{Q}_t^T \cdots \textbf{Q}_{   t-k}^T \odot \hat{\mathbf{x}}_0^i \overline{\mathbf{Q}}_{t-k}}{\hat{\mathbf{x}}^{i}_0 \overline{\mathbf{Q}}_t (\mathbf{x}_t^i)^T}\right),
\end{align}

\noindent where $k$ is the number of skipping steps.

Then we introduce the application of Monte-Carlo dropout within the generative process, a technique designed to enhance prediction uncertainty in neural networks. Specifically, we utilize dropout not only to prevent overfitting during the training of our denoising network, but also maintain its activation in the inference stage. By keeping dropout enabled and running multiple forward passes (Monte-Carlo samples) during inference, we generate a prediction distribution for each input, as opposed to a single-point estimation. To improve uncertainty estimation, we aggregate the predictions by taking a mean pooling over all output logits corresponding to the same input. This operation leads to the predicted logits that perform reduced estimation bias, and their normalized probabilities can more accurately reflect the actual distribution. Therefore, we can leverage Monte-Carlo dropout to enhance the generative process toward more reliable samplings.

\subsection*{Mask-prior-guided denoising network}
In diffusion model applications, the denoising network plays a crucial role in generation performance. We develop a mask-prior-guided denoising network, integrating both structural information and residue interactions for enhanced protein sequence prediction. Our denoising network architecture encompasses a structure-based sequence predictor, a pre-trained mask sequence designer, and a mask ratio adapter.

\noindent \textbf{Structure-based sequence predictor.} We adopt an equivariant graph neural network (EGNN) with a global-aware module as the structure-based sequence predictor, which generates a full AA sequence from the backbone structure. EGNN is a type of graph neural network that satisfies equivariance operations for the special Euclidean group SE(3). It preserves geometric and spatial relationships of 3D coordinates within the message-passing framework. Given a noise residue graph , we use $\mathbf{H} = [\mathbf{h}_1, \mathbf{h}_2, \cdots, \mathbf{h}_N]$ to denote the initial node embeddings, which are derived from the noisy AA types and geometric properties. The coordinates of each node is represented as $\mathbf{X}^\mathrm{pos} = [\mathbf{x}_1^\mathrm{pos}, \mathbf{x}_2^\mathrm{pos}, \cdots \mathbf{x}_N^\mathrm{pos}]$, while the edge feature are denoted as $\mathbf{E} = [\mathbf{e}_{1}, \mathbf{e}_{2}, \cdots \mathbf{e}_M]$. In this setting, EGNN consists of a stack  of equivariant graph convolutional layers (EGCL) for the node and edge information propagation, which are defined as:

\begin{align}
    \mathbf{e}_{ij}^{(l+1)} &= \phi_{e}(\mathbf{h}_i^{(l)}, \mathbf{h}_j^{(l)}, \Vert \mathbf{x}_i^{(l)} - \mathbf{x}_j^{(l)} \Vert^2, \mathbf{e}_{ij}^{(l)}), \\
    \hat{\mathbf{h}}_i^{(l+1)} &= \phi_{h}(\mathbf{h}_i^{(l)}, \sum_{j \in \mathcal{N}(i)} w_{ij} \mathbf{e}_{ij}^{(l + 1)}), \\
    \mathbf{x}_i^{(l+1)} &= \mathbf{x}_i^{(l)} + \frac{1}{N_i} \sum_{j \in \mathcal{N}(i)} (\mathbf{x}_i^{(l)} - \mathbf{x}_j^{(l)}) \phi_x(\mathbf{e}_{ij}^{(l + 1)}),
\end{align}

\noindent where $l$ denotes the $l$-th EGCL layer, $\mathbf{x}_i^{(0)} = \mathbf{x}_i^\mathrm{pos}$, and $w_{ij} = \mathrm{sigmoid}(\phi_w(\mathbf{e}_{ij}^{(l+1)})))$ is a soft estimated weight assigned to the specific edge representation. All components ($\phi_e$, $\phi_h$, $\phi_x$, $\phi_w$) are learnable and parametrized by fully connected neural networks. In the information propagation, EGNN achieves equivariance to translations and rotations on the node coordinates $\mathbf{X}^\mathrm{pos}$, and preserves invariant to group transformations on the node features $\mathbf{H}$ and edge features $\mathbf{E}$. 

However, the vanilla EGNN only considers local neighbour aggregation while neglecting the global context. Some recent studies \cite{tan2023global, gao2023pifold} have demonstrated the importance of global information in protein design. Therefore, we introduce a global-aware module in the EGCL layer, which incorporates the global pooling vector into the update of node representations:

\begin{align}
    \mathbf{m}^{(l+1)} &= \mathrm{MeanPool}(\{\hat{\mathbf{h}}_i^{(l+1)}\}_{i \in \mathcal{G}}), \\
    \mathbf{h}_i^{(l+1)} &= \hat{\mathbf{h}}_i^{(l+1)} \odot \mathrm{sigmoid}(\phi_m(\mathbf{m}^{(l+1)}, \hat{\mathbf{h}}_i^{(l+1)})),
\end{align}

\noindent where $\mathrm{MeanPool}(\cdot)$ is the mean pooling operation over all nodes within a residue graph. The global-aware module effectively integrates global context into modelling and only increases a linear computational cost. To predict the probabilities of residue types, the node representations from the last EGCL layer are fed into a fully connected classification layer with softmax function, which is defined as:

\begin{align}
    \textbf{p}_i^\mathrm{b} = \mathrm{softmax}(\textbf{l}_i^\mathrm{b}), \quad \textbf{l}_i^\mathrm{b} = \mathbf{h}_{i}^{(L)} \mathbf{W}_\mathrm{o} + \mathbf{b}_\mathrm{o},
\end{align}

\noindent where $\mathbf{W}_\mathrm{o} \in \mathbb{R}^{D_h \times 20}$ and $\mathbf{b}_\mathrm{o} \in \mathbb{R}^{1 \times 20}$ are learnable weight matrix and bias vector. 

\noindent \textbf{Low-confidence residue selection and mask ratio adapter.} As previously mentioned, structural information alone can sometimes be insufficient to determine all residue identities. Certain flexible regions display a weaker correlation with the backbone structure but are strongly influenced by their sequential context. To enhance the denoising network's performance, we introduce a masked sequence designer module. This module refines the residues identified with low confidence in the base sequence predictor. We adopt an entropy-based residue selection strategy, as proposed by Zhou et al. (2023) \cite{zhou2023prorefiner}, to identify these low-confidence residues. The entropy for the $i$-th residue of the probability distribution $\mathbf{p}_i^\mathrm{b}$ is calculated as:

\begin{align}
    \mathrm{ent}_i^\mathrm{b} = -\sum_j \text{p}^\mathrm{b}_{ij} \log(\text{p}^\mathrm{b}_{ij}).
\end{align}

Given that entropy quantifies the uncertainty in a probability distribution, it can be utilized to locate the low-confidence predicted residues. Consequently, residues with the most entropy are masked, while the rest remain in a sequential context. The masked sequence designer aims to reconstruct the entire sequence by using the masked partial sequence in combination with the backbone structure. In addition, to account for the varying noise levels of the input sequence in diffusion models, we design a simple mask ratio adapter to dynamically determine the entropy mask percentage at different denoising steps:

\begin{align}
    \text{mr}_t = \sin\left(\frac{\pi}{2} \beta_t\sigma \right) + m,
\end{align}

\noindent where $\beta_t \in [0, 1]$ represents the noise weight at step $t$ derived from the noise schedule, and $\sigma$ and $m$ are predefined deviation and minimum mask ratio, respectively. With the increase of $\beta_t$, the mask ratio is proportional to its time step.

\noindent \textbf{Mask-prior pre-training.} To incorporate prior knowledge of sequential context, we pre-train the masked sequence designer by applying the masked language modelling objective proposed in BERT \cite{devlin-etal-2019-bert}. It is important to clarify that we use the same training data in the diffusion models for pre-training purposes, in order to avoid any information leakage from external sources. In this process, we randomly sample a proportion of residues in the native AA sequences, and replace them with the masking procedures: (i) masking 80\% of the selected residues using a special MASK type, (ii) replacing 10\% of the selected residues with other random residue types, and (iii) keeping the remaining 10\% residues unchanged. Subsequently, we input the partially masked sequences, along with structural information, into the masked sequence designer. The objective of the pre-training stage is to predict the original residue types from the masked residue representations using a cross-entropy loss function.

\noindent \textbf{Masked sequence designer.} We use an invariant point attention (IPA) network as the masked sequence designer. IPA is a geometry-aware attention mechanism designed to facilitate the fusion of residue representations and spatial relationships, enhancing the structure generation within AlphaFold2 \cite{jumper2021highly}. In this study, we repurpose the IPA module to refine low-confidence residues in the base sequence predictor. Given a mask AA sequence, we denote its residue representation as $\mathbf{S} = [\mathbf{s}_1, \mathbf{s}_2, \cdots, \mathbf{s}_N]$, which is derived from the residue types and positional encoding. To incorporate geometric information, as with the IPA implementation in Frame2seq \cite{akpinaroglu2023structure}, we construct a pairwise distance representation $\mathbf{Z} = \{\mathbf{z}_{ij} \in \mathbb{R}^{1 \times d_z} \mid 1 \leq i \leq N, 1 \leq j \leq N\}$ and rigid coordinate frames $\mathcal{T} = \{T_i := (\mathbf{R}_i \in \mathbb{R}^{3 \times 3}, \mathbf{t}_i \in \mathbb{R}^3) \mid 1 \leq i \leq N\}$. The pairwise representation $\mathbf{Z}$ is obtained by calculating inter-residue spatial distances and relative sequence positions. \textcolor{myblue}{The rigid coordinate frames are constructed from the coordinates of backbone atoms using a Gram-Schmidt process, providing a consistent local reference for ensuring the invariance of IPA to global Euclidean transformations.} Subsequently, we take the residue representation, pairwise distance representation and rigid coordinate frames as inputs, and feed them into a stack of IPA layers for representation learning, which is defined as:

\begin{align}
    \mathbf{S}^{(l+1)}, \mathbf{Z}^{(l+1)} = \mathrm{IPA}(\mathbf{S}^{(l)}, \mathbf{Z}^{(l)}, \mathcal{T}).
\end{align}

The IPA network follows the self-attention mechanism. However, it enhances the general attention queries, keys, and values by incorporating 3D points that are generated in the rigid coordinate frame of each residue. This operation ensures that the updated residue and pair representations remain invariant by global rotations and translations. More details on the IPA feature construction and algorithm implementation are provided in Supplementary S6. For the $i$-th residue, the predicted probability distribution and entropy in the masked sequence designer are calculated as:

\begin{align}
    \textbf{p}_i^\mathrm{m} = \mathrm{softmax}(\textbf{l}_i^\mathrm{m}), \quad \textbf{l}_i^\mathrm{m} = \mathbf{h}_{i}^{(L)} \mathbf{W}_\mathrm{m} + \mathbf{b}_\mathrm{m},
\end{align}

\begin{align}
    \mathrm{ent}_i^\mathrm{m} = -\sum_j \text{p}^\mathrm{m}_{ij} \log(\text{p}^\mathrm{m}_{ij}),
\end{align}

\noindent where $\mathbf{W}_\mathrm{m} \in \mathbb{R}^{D_s \times 20}$ and $\mathbf{b}_\mathrm{m} \in \mathbb{R}^{1 \times 20}$ are learnable weight matrix and bias vector. The training objective is to jointly minimize the cross-entropy losses for both the base sequence predictor and masked sequence designer. In the inference stage, we calculate the final predicted probability by weighting the output logits based on their entropy as follows:

\begin{align}
    \mathbf{l}_i^{\text{f}} = \frac{\mathrm{exp}({-\mathrm{ent}_i^\mathrm{b}})}{\mathrm{exp}({-\mathrm{ent}_i^\mathrm{b}}) + \mathrm{exp}({-\mathrm{ent}_i^\mathrm{m}})} \textbf{l}_i^\mathrm{b} + \frac{\mathrm{exp}({-\mathrm{ent}_i^\mathrm{m}})}{\mathrm{exp}({-\mathrm{ent}_i^\mathrm{b}}) + \mathrm{exp}({-\mathrm{ent}_i^\mathrm{m}})} \textbf{l}_i^\mathrm{m}.
\end{align}

\begin{align}
    \mathbf{p}_i^{\text{f}} = \mathrm{softmax}(\textbf{l}_i^{\text{f}}).
\end{align}

By incorporating the mask-prior denoising network into the discrete denoising diffusion process, our framework enhances the denoising trajectories, leading to more accurate predictions of protein sequences.

\subsection*{Experimental setting}
\noindent \textcolor{myblue}{\textbf{Primary datasets.}  
We evaluate MapDiff on experimentally validated protein structures curated from well-established databases. The CATH database \cite{orengo1997cath} is widely used in inverse folding research, enabling fair comparisons across different methodologies. It classifies proteins into hierarchical levels based on class, architecture, topology, and homologous superfamily, with filtering to reduce redundancy and ensure structural diversity. Following previous studies \cite{ingraham2019generative, hsu2022learning, gao2023pifold}, proteins are partitioned based on their CATH topology classification codes, ensuring that the training, validation, and test sets contain non-overlapping topologies. This partitioning strategy provides a robust evaluation of the model’s generalization to unseen proteins. For CATH 4.2, the dataset consists of 18,024 structures for training, 608 for validation, and 1,120 for testing. Similarly, in CATH 4.3, we follow the topology classification approach in ESM-IF \cite{hsu2022learning}, resulting in 16,630 proteins for training, 1,516 for validation, and 1,864 for testing. By including both CATH 4.2 and CATH 4.3, we assess the stability of model performance across dataset versions, ensuring robustness to updates in protein structure databases.}

\noindent \textcolor{myblue}{\textbf{Zero-shot generalization datasets.} 
To further assess MapDiff’s zero-shot generalization ability, we evaluate it on the two independent TS50 and PDB2022 datasets. TS50 \cite{li2014direct} is a commonly used benchmark for protein sequence design, consisting of 50 diverse protein chains covering different structural classes. PDB2022 includes single-chain structures published in the Protein Data Bank (PDB) \cite{berman2000protein} between 5 January 2022 and 26 October 2022, curated by Zhou et al. (2022) \cite{zhou2023prorefiner}, with protein length $\leq$ 500 and resolution $\leq$ 2.5 Å. This dataset consists of 1,975 proteins published after those in the CATH dataset, ensuring a strict time-based test ``split'' to evaluate real-world temporal generalization. Both datasets are entirely separate from the CATH-derived training set, minimizing data leakage and providing a robust evaluation of structural and temporal generalization.}

\noindent \textbf{Baselines.} We compare MapDiff with recent deep graph models for inverse protein folding, including StructGNN \cite{ingraham2019generative}, GraphTrans \cite{ingraham2019generative}, GVP \cite{jing2020learning}, AlphaDesign \cite{gao2022alphadesign}, ProteinMPNN \cite{dauparas2022robust}, PiFold \cite{gao2023pifold}, LM-Design \cite{lmdesign} and GRADE-IF \cite{dauparas2022robust}. To ensure a reliable and fair comparison, we reproduce the open-source and four most state-of-the-art baselines (ProteinMPNN, PiFold, LM-Design, and GRADE-IF) under identical settings in our experiments. ProteinMPNN uses a message-passing neural network to encode structure features, and a random decoding scheme to generate protein sequences. PiFold introduces a residue featurizer to extract distance, angle, and direction features. It proposes a PiGNN encoder to learn expressive residue representations, enabling the generation of protein sequences in a one-shot manner. LM-Design uses structure-based models as encoders and incorporates the protein language model ESM as a protein designer to refine the generated sequences. GRADE-IF employs EGNN to learn residue representations from protein structures, and adopts the graph denoising diffusion model to iteratively generate feasible sequences. All baselines are implemented following the default hyperparameter settings in their original papers.

\noindent \textbf{Implementation setup.} MapDiff is implemented in Python 3.8 and PyTorch 1.13.1 \cite{paszke2017automatic}, along with functions from BioPython 1.81 \cite{cock2009biopython}, PyG 2.4.0 \cite{pyg}, Scikit-learn 1.0.2 \cite{pedregosa2011scikit}, NumPy 1.22.3 \cite{harris2020array}, and RDKit 2023.3.3 \cite{rdk}. It consists of two training stages: mask-prior pre-training and denoising diffusion model training, both of which use the same CATH 4.2/4.3 training set. The batch size is set to 8, and the models are trained up to 200 epochs in pre-training and 100 epochs in denoising training. We employ the Adam optimizer with a one-cycle scheduler for parameter optimization, setting the peak learning rate to 5e-4. In the denoising network, the structure-based sequence predictor consists of six global-aware EGCL layers, each with 128 hidden dimensions. Meanwhile, the masked sequence designer stacks six layers of IPA, each with 128 hidden dimensions and 4 attention heads. The dropout rate is set to 0.2 in both EGCL and IPA layers. A cosine schedule is applied to control the noise weight at each time steps, with a total of 500 time steps. During sampling inference, the skip steps for DDIM are configured to 100, and the Monte-Carlo forward passes are set to 50. For the mask ratio adaptor, we set the minimum mask ratio to 0.4 and the deviation to 0.2. All experiments are conducted on a single Tesla A100 GPU. Following the regular evaluation in deep learning, the best-performing model is selected based on the epoch that provides the highest recovery on the validation set. \textcolor{myblue_2}{After that}, this selected model is subsequently used to evaluate performance on the test set. \textcolor{myred_2}{For the foldability analysis, we apply a single AlphaFold2 pTM model (i.e., model\_1\_ptm) with three recycles to balance accuracy and computational efficiency. Multiple sequence alignment (MSA) information is generated for each sequence using the MMSeqs2 \cite{steinegger2017mmseqs2, mirdita2019mmseqs2} server provided by ColabFold \cite{mirdita2022colabfold}}. Additionally, we provided the algorithm details for the training and sampling inference in Supplementary S5, and the scalability study \textcolor{myblue}{in Supplementary S8 and Supplementary Fig. 4}.

\section*{Data availability}
\textcolor{myred_2}{The experimental data used in this work is available at \url{https://github.com/peizhenbai/MapDiff/tree/main/data}. All data were publicly collected from the following resources.} The CATH 4.2 dataset can be found at \url{https://github.com/dauparas/ProteinMPNN}; the CATH 4.3 dataset can be found at \url{https://github.com/BytedProtein/ByProt}; the PDB2022 dataset can be found at \url{https://github.com/veghen/ProRefiner} and the TS50 dataset can be found at \url{https://github.com/A4Bio/PiFold}. The protein structure data is obtained from Protein Data Bank \url{https://www.rcsb.org/} with the corresponding PDB IDs.

\section*{Code availability}
The source code and implementation details of MapDiff are available at both GitHub repository (\url{https://github.com/peizhenbai/MapDiff}) and CodeOcean capsule (\url{https://doi.org/10.24433/CO.3441652.v1}). The code is also archived on Zenodo (\url{https://doi.org/10.5281/zenodo.15162932}) \cite{bai2025mapdiff_zeno}.

\section*{Acknowledgements}
\textcolor{myred_2}{We are grateful to T. Ucar, X. Song, and S. Zhou for their invaluable suggestions on the work. P.B. received the Faculty of Engineering Research Scholarship at the University of Sheffield.}

\section*{Author contributions}
\textcolor{myred_2}{P.B. developed the models and conceived and designed the experiments under the guidance of L.D.M, R.C.W, O.R., and H.L. F.M., X.L., and P.B. contributed to analysis tools, performed the experiments, and conducted method comparisons. All authors contributed to analyze the data and write the paper.}

\section*{Competing interests}
The authors declare no competing interests.


\clearpage
\section*{\Large{Supplementary Material}}

\subsection*{\textcolor{red}{S1. Closest training structures and sequences}}
\textcolor{red}{In Supplementary Fig. \ref{fig:tm_score}, we present the structure ID with the closest TM score and the sequence ID with the highest bit-score (calculated by BLAST \cite{altschul1990basic}) in the training set, compared to the three sequences (PDB ID: 1NI8, 2HKY and 2P0X) generated by MapDiff and their refolded structures predicted by AlphaFold2. The TM-score evaluates structural similarity based on 3D atomic coordinates, while the bit-score quantifies sequence similarity by assessing alignment quality.}

\textcolor{red}{There are two observations. Firstly, we observe a weak positive correlation between similarity in the training set and evaluation set. In Supplementary Fig. \ref{fig:tm_score}a, the structure most similar to 1NI8 has the highest TM-score (0.57), which results in the highest TM-score (0.73) for the AlphaFold2-refolded MapDiff structure. For the bit-score in Supplementary Fig. \ref{fig:tm_score}b, a higher value for the closest sequence in the training set corresponds to a higher alignment score for the predicted sequence.}

\textcolor{red}{Secondly, the training set has no samples with high similarity (TM-score > 0.7 or bit-score > 50) relative to the test cases. This is due to the high structural diversity of proteins in CATH and the train/validation/test split based on their topology classification codes, which ensures no overlap in topology between different sets.}

\renewcommand*{\figurename}{Supplementary Figure}
\setcounter{figure}{0}
\begin{figure}[htbp]
\centering
    \includegraphics[width=1\textwidth]{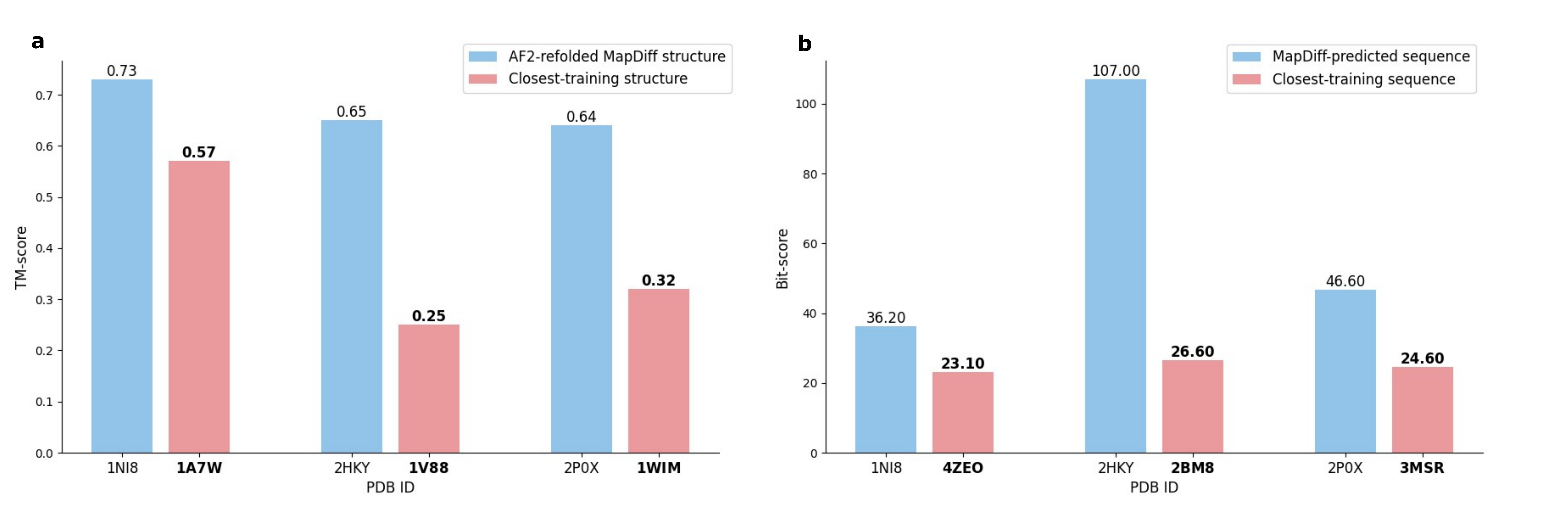}
    \caption{\textbf{\textcolor{red}{Comparison of the closest structural and sequence matches in the training set for the three selected test cases (PDB ID: 1NI8, 2HKY, and 2P0X).}} \textcolor{red}{(a) TM-score comparison between the AlphaFold2-refolded MapDiff structure and the closest training structures (PDB ID: 1A7W, 1V88 and 1WIM), relative to the test structures. (b) Bit-score comparison between the MapDiff-predicted sequence and the closest training sequences (PDB ID: 4ZEO, 2BM8 and 3MSR), relative to the test sequence.}}
    \label{fig:tm_score}
\end{figure}

\begin{figure}[htbp]
\centering
    \includegraphics[width=1\textwidth]{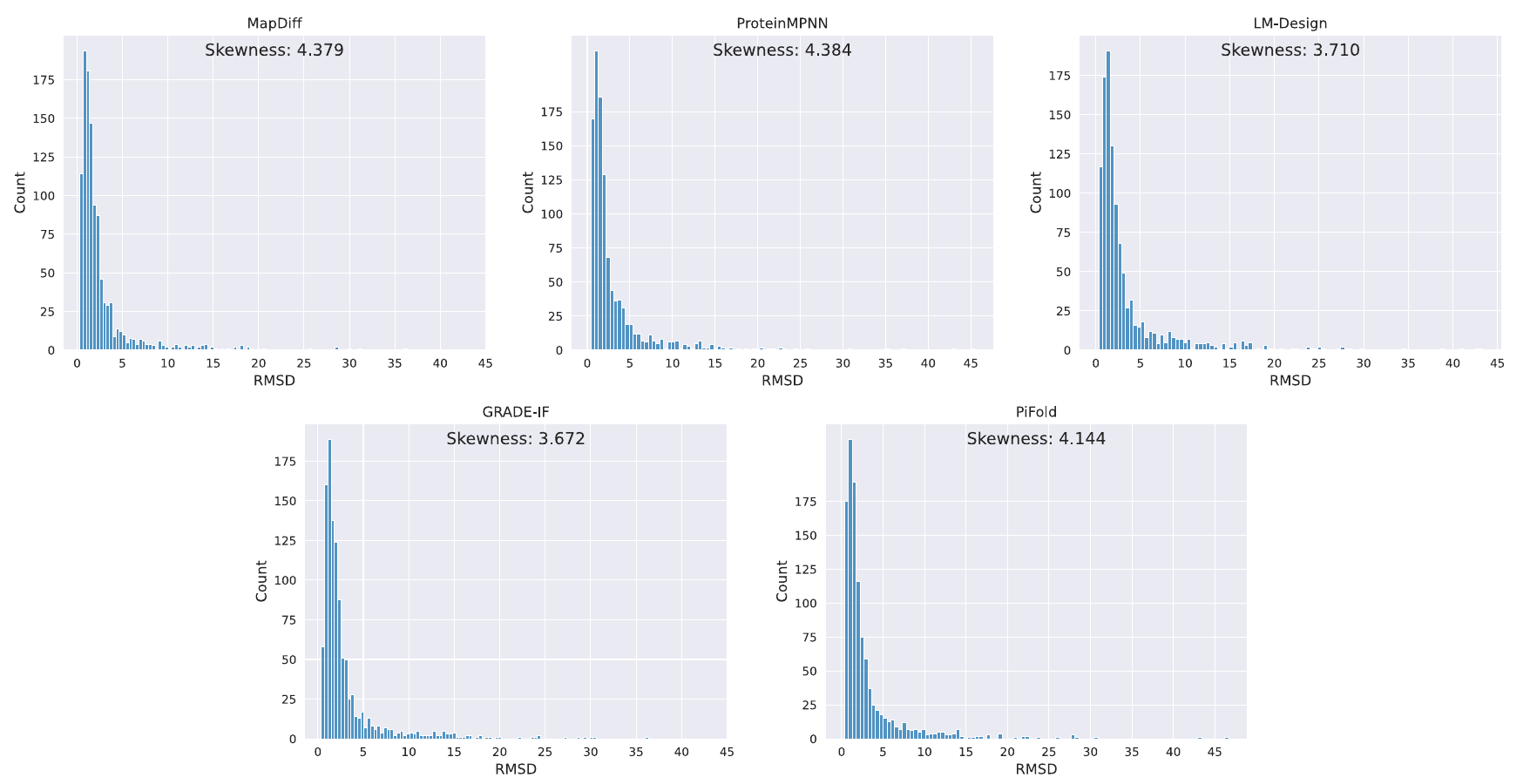}
    \caption{\textcolor{red}{Histogram and skewness of RMSD distributions for generated sequences on the CATH 4.2 test set ($\textit{n}$ = 1,120), refolded using AlphaFold2.}}
    \label{fig:rmsd_his_su}
\end{figure}

\subsection*{\textcolor{red}{S2. Histogram of RMSD distributions}}

\textcolor{red}{For the foldability measurement using AlphaFold2, the standard deviation of RMSD is consistently larger than the mean across all models. This occurs because the RMSD values follow a right-skewed distribution \cite{log_normal} rather than a normal (Gaussian) distribution. To confirm this, we present the histogram and skewness of RMSD for the five models in Supplementary Fig. \ref{fig:rmsd_his_su}. It shows that each distribution has a few extremely high values (RMSD > 10) that pull the tail to the right, thus increasing both the mean and standard deviation. The outliers arise from inherent flaws in the generated sequences or biases in AlphaFold2 during the refolding process. Skewness measures the asymmetry of a distribution. It quantifies whether a distribution has more values concentrated on one side of the mean and how extreme the tail is on the other side. We can see that the skewness of all models' RMSD is greater than 3.5, which indicates that they have a heavy right tail. As a result, the magnitude of the standard deviation is larger than the mean in RMSD distributions.}

\subsection*{S3. Discrete posterior distribution}
We present the derivations of the discrete posterior distribution in the reverse denoising process. To clarify the derivation in equation (8) of the main text, we remove the residue index $i$ and assume $\mathbf{x}_{t} = \mathbf{x}_{t}^{i}$, as it is not relevant to the theoretical derivation.

\noindent \textbf{Propsition 1.} For $q({\mathbf{x}_{t-1} \mid \mathbf{x}_{t}, \mathbf{x}_{0}})$ defined in equation (3) of the main text, we have:

\begin{equation}
q({\mathbf{x}_{t-1} \mid \mathbf{x}_{t}, \mathbf{x}_{0}}) \propto \mathbf{x}_t \mathbf{Q}_t^T \odot \mathbf{x}_0 \overline{\mathbf{Q}}_{t-1},
\end{equation}

\noindent \textit{Proof.} By Bayesian theorem, the distribution can be written as:

\begin{equation}
q(\mathbf{x}_{t-1} \mid \mathbf{x}_{t}, \mathbf{x}_{0}) \propto q(\mathbf{x}_{t} \mid \mathbf{x}_{t-1}, \mathbf{x}_{0}) q({\mathbf{x}_{t-1} \mid \mathbf{x}_{0}}).
\end{equation}

Due to the Markov property, $q({\mathbf{x}_{t} \mid \mathbf{x}_{t-1}, \mathbf{x}_{0}})$ is equivalent to $q(\mathbf{x}_{t} \mid \mathbf{x}_{t-1})$. Similarly, we have: 

\begin{equation}
q(\mathbf{x}_{t} \mid \mathbf{x}_{t-1}) \propto q(\mathbf{x}_{t-1} \mid \mathbf{x}_{t}) q(\mathbf{x}_{t}).
\end{equation}

 Given that $q(\mathbf{x}_{t})$ is an independent distribution, it can be viewed as a normalizing constant. By the definition of the discrete transition matrix $\mathbf{Q}_t$, we can derive:

\begin{align}
    q(\mathbf{x}_{t-1} \mid \mathbf{x}_{0}) &= \mathrm{Cat}(\mathbf{x}_{t-1}; \mathbf{p} = \mathbf{x}_{0}\overline{\mathbf{Q}}_{t-1}), \\
    q(\mathbf{x}_{t-1} \mid \mathbf{x}_{t}) &= \mathrm{Cat}(\mathbf{x}_{t-1}; \mathbf{p} = \mathbf{x}_t \mathbf{Q}_t^T).
\end{align}

\noindent By integrating the above terms, we obtain $q({\mathbf{x}_{t-1} \mid \mathbf{x}_{t}, \mathbf{x}_{0}}) \propto \mathbf{x}_t \mathbf{Q}_t^T \odot \mathbf{x}_0 \overline{\mathbf{Q}}_{t-1}$ as intended.

\noindent \textbf{Propsition 2.} For $q({\mathbf{x}_{t-1} \mid \mathbf{x}_{t}, \hat{\mathbf{x}}_{0}})$ defined in equation (8) of the main text, we have:

\begin{align}
    q({\mathbf{x}_{t-1} \mid \mathbf{x}_{t}, \hat{\mathbf{x}}_{0}}) 
    &= \mathrm{Cat}\left(\mathbf{x}_{t-1}; \mathbf{p} = \frac{\textbf{x}_t \textbf{Q}_t^T \odot \hat{\mathbf{x}}_0 \overline{\mathbf{Q}}_{t-1}}{\hat{\mathbf{x}}_0 \overline{\mathbf{Q}}_t \mathbf{x}_t^T}\right).
\end{align}

\noindent \textit{Proof}. By Bayesian theorem, we write the distribution as:
\begin{align}
    q({\mathbf{x}_{t-1} \mid \mathbf{x}_{t}, \hat{\mathbf{x}}_{0}}) &= \frac{q(\mathbf{x}_t \mid \mathbf{x}_{t-1}, \hat{\mathbf{x}}_{0}) q(\mathbf{x}_{t-1} \mid \hat{\mathbf{x}}_{0})}{q(\mathbf{x}_{t} \mid \hat{\mathbf{x}}_{0})}.
    \label{post_infer}
\end{align}

According to the definition of the reverse denoising process, both random variables $X_0 = \hat{\mathbf{x}}_0$ and $X_t = \mathbf{x}_t$ are observed. Here, $\mathbf{x}_t$ represents the discrete noise state at step $t$, while $\hat{\mathbf{x}}_0$ denotes the predicted clean state by the denoising neural network. Therefore, we can rewrite equation (\ref{post_infer}) in terms of the observed variables:

\begin{align}
    q({X_{t-1} \mid X_t = \mathbf{x}_{t}, X_0 = \hat{\mathbf{x}}_{0}}) &= \frac{q(X_t = \mathbf{x}_t \mid \mathbf{x}_{t-1}, X_0 = \hat{\mathbf{x}}_{0}) q(\mathbf{x}_{t-1} \mid X_0 = \hat{\mathbf{x}}_{0})}{q(X_t = \mathbf{x}_{t} \mid X_0 = \hat{\mathbf{x}}_{0})}.
    \label{post_infer2}
\end{align}

Due to the Markov property, we have:

\begin{align}
    q({X_{t-1} \mid X_t = \mathbf{x}_{t}, X_0 = \hat{\mathbf{x}}_{0}}) &= \frac{q(X_t = \mathbf{x}_t \mid X_{t-1}) q(\mathbf{x}_{t-1} \mid X_0 = \hat{\mathbf{x}}_{0})}{q(X_t = \mathbf{x}_{t} \mid X_0 = \hat{\mathbf{x}}_{0})}.
    \label{post_prob}
\end{align}

Now we can individually define each term in equation (\ref{post_prob}). Regarding the distribution $q(X_t = \mathbf{x}_t \mid X_{t-1})$, it is important to note that the value of $X_{t-1}$ is not observed. Consequently, it is unfeasible to compute a determined result for the conditional distribution. Instead, we need to enumerate all possible $\mathbf{x}_{t-1}$ to compute the probability distribution over $\mathbf{x}_t$, defined as follows:

\begin{align}
    q(X_t = \mathbf{x}_t \mid X_{t-1}) &= \mathbf{x}_t[\mathbf{I}_k \mathbf{Q}_t]^T, \\ 
    & = \mathbf{x}_t \mathbf{Q}_t^T.
\end{align}

For the two terms $q(\mathbf{x}_{t-1} \mid X_0 = \hat{\mathbf{x}}_{0})$ and $q(X_t = \mathbf{x}_{t} \mid X_0 = \hat{\mathbf{x}}_{0})$, we can extend them following the definition of the forward diffusion process:

\begin{align}
    &q(\mathbf{x}_{t-1} \mid X_0 = \hat{\mathbf{x}}_{0}) = \hat{\mathbf{x}}_0 \overline{\mathbf{Q}}_{t-1}, \\
    &q(X_t = \mathbf{x}_{t} \mid X_0 = \hat{\mathbf{x}}_{0}) = \hat{\mathbf{x}}_0 \overline{\mathbf{Q}}_t \mathbf{x}_t^T.
\end{align}

By putting them all together, we obtain:

\begin{align}
    q({\mathbf{x}_{t-1} \mid \mathbf{x}_{t}, \hat{\mathbf{x}}_{0}}) 
    &= \mathrm{Cat}\left(\mathbf{x}_{t-1}; \mathbf{p} = \frac{\textbf{x}_t \textbf{Q}_t^T \odot \hat{\mathbf{x}}_0 \overline{\mathbf{Q}}_{t-1}}{\hat{\mathbf{x}}_0 \overline{\mathbf{Q}}_t \mathbf{x}_t^T}\right).
\end{align}

\begin{algorithm}[h]
   \caption{Denoising training algorithm of MapDiff}
   \label{alg:training}
\begin{algorithmic}[1]
\STATE{\bf Input:} residue graph $\mathcal{G} = (\mathbf{X}, \mathbf{A}, \mathbf{E})$, where $\mathbf{X} = [\mathbf{X}^\mathrm{aa}_0, \mathbf{X}^\mathrm{pos}, \mathbf{X}^\mathrm{prop}]$, pairwise representation $\mathbf{Z}$,  rigid coordinate frames $\mathcal{T}$, denoising time steps $T$
\STATE{\bf Initial:} base sequence predictor $h_\phi$, masked sequence designer $f_\theta$
    \WHILE{$\phi,\theta$ have not converged}
    \STATE{\bf First step: base sequence prediction}
    \STATE Sample time step $t \sim \mathcal{U}(1,T)$ and noisy sequence $\mathbf{X}_t^\mathrm{aa} \sim q(\mathbf{X}_t^\mathrm{aa} \mid \mathbf{X}_0^\mathrm{aa})$
    \STATE Predict $p_\mathrm{b}(\hat{\mathbf{X}}^\mathrm{aa}_0)$ = $h_\phi(\mathbf{X}_t^\mathrm{aa}, \mathbf{X}^\mathrm{pos}, \mathbf{X}^\mathrm{prop}, \mathbf{E}, t)$ \hfill\COMMENT{Global-aware EGNN}
    \STATE Compute base cross-entropy loss $L_\mathrm{b} = L_\mathrm{CE}(p(\hat{\mathbf{X}}^\mathrm{aa}_0), \mathbf{X}^\mathrm{aa}_0)$
    \STATE{\bf Second step: masked sequence refinement}
    \STATE Compute base entropy $\{\mathrm{ent}^b_1, \cdots \mathrm{ent}^b_N\} = E(p(\hat{\mathbf{X}}^\mathrm{aa}_0))$
    \STATE Compute mask ratio $\mathrm{mr}_t =  \sin\left(\frac{\pi}{2} \beta_t \sigma \right) + m$
    \STATE Generate entropy-based masked sequence and residue representations $\mathbf{S} = [\mathbf{s}_1, \mathbf{s}_2, \cdots, \mathbf{s}_N]$
    \STATE Predict $p_\mathrm{m}(\{\hat{\mathbf{X}}^\mathrm{aa}_0\}_\mathrm{mask})$ = $f_\theta(\mathbf{S}, \mathbf{Z}, \mathcal{T})$ \hfill\COMMENT{IPA network with mask pre-training}
    \STATE Compute mask cross-entropy loss $L_\mathrm{m} = L_\mathrm{CE}(p(\{\hat{\mathbf{X}}^\mathrm{aa}_0\}_\mathrm{mask}), \{\mathbf{X}^\mathrm{aa}_0\}_\mathrm{mask})$
    \STATE Compute total loss $L = L_\mathrm{b} + L_\mathrm{m}$
    \STATE Update $\phi, \theta \leftarrow \mathrm{optimizer} (L, \phi, \theta)$
\ENDWHILE
\STATE \bf{return} $h_\phi$, $f_\theta$
\end{algorithmic}
\label{training_alg}
\end{algorithm}

\subsection*{S4. Residue graph feature construction}
The protein is represented as a residue graph $\mathcal{G} = (\mathbf{X}, \mathbf{A}, \mathbf{E})$ to reflect its geometric structure and topological relationships. The node feature $\mathbf{X} \in \mathbb{R}^{N \times 42}$ is decomposed into three components: $\mathbf{X}^\mathrm{aa} \in \mathbb{R}^{N \times 20}, \mathbf{X}^\mathrm{pos} \in \mathbb{R}^{N \times 3}$ and $\mathbf{X}^\mathrm{prop} \in \mathbb{R}^{N \times 19}$. Each row of $\mathbf{X}^\mathrm{aa}$ represents the one-hot encoded vector that indicates the amino acid type of a residue. $\mathbf{X}^\mathrm{pos}$ denotes the spatial backbone coordinates, specifically represented by the backbone $C_{\alpha}$ atoms. Following Yi et al. (2023) \cite{yi2024graph} and Ganea et al. (2022) \cite{ganea2022independent}, we describe the structural and geometric properties of the residues by $\mathbf{X}^\mathrm{prop}$, including the solvent-accessible surface area (SASA), crystallographic B-factor, surface-aware node features, protein secondary structure and backbone dihedral angles. Protein secondary structure reflects the local folding patterns of AA residues in a protein chain. We utilize the DSSP \cite{kabsch1983dictionary} algorithm to calculate the secondary structures for each residue and represent them by one-hot encoding. We derive the sin and cos values for the backbone dihedral angles $\phi$ and $\psi$, denoting the spatial arrangement within the backbone atoms. The surface-aware node features are the weighted average distance of each residue node to its one-hop neighbors, defined as follows:

\begin{equation}
    \rho_i(\mathbf{x}_i; \lambda) = \frac{\|\sum_{i' \in \mathcal{N}_i} w_{i,i', \lambda} (\mathbf{x}_i^\mathrm{pos} - \mathbf{x}_{i'}^\mathrm{pos}) \| } {\sum_{i' \in \mathcal{N}_i} w_{i,i', \lambda} \|\mathbf{x}_i^\mathrm{pos} - \mathbf{x}_{i'}^\mathrm{pos}\|}, \quad \text{where\ } w_{i,i', \lambda} = \frac{\exp(- ||\mathbf{x}_i^\mathrm{pos} - \mathbf{x}_{i'}^\mathrm{pos}||^2 / \lambda)}{\sum_{j \in \mathcal{N}_i} \exp(- ||\mathbf{x}_i^\mathrm{pos} - \mathbf{x}_{j}^\mathrm{pos}||^2 / \lambda)},
    \label{surface_feat}
\end{equation}

\noindent where $\lambda \in \{1., 2., 5., 10., 30.\}$.

The edge feature $\mathbf{E} \in \mathbb{R}^{M \times 93}$ describes connections between pairs of residue nodes, including the relative spatial distances, local spatial positions and relative sequential positions. The relative spatial distance is defined as the Euclidean distance between the $C_\alpha$ coordinates of two residues. This distance is then projected into a 15-dimensional kernel space using a radial basis function (RBF). In addition, a binary contact signal \cite{ingraham2019generative} is used to indicate if the spatial distance between two residues is less than 8 \AA. The local spatial positions \cite{ganea2022independent} have 12 dimensions and are created from a local coordinate system. They represent the relative positions and local orientations among the backbone atoms of residues. Next, the relative sequential position is defined as a 65-dimensional one-hot feature, based on the difference in the sequential indices along the AA chain.

For the IPA network, we further construct a pairwise distance representation $\mathbf{Z} = \{\mathbf{z}_{ij} \in \mathbb{R}^{1 \times d_z} \mid 1 \leq i \leq N, 1 \leq j \leq N\}$ and rigid coordinate frames $\mathcal{T} = \{T_i := (\mathbf{R}_i \in \mathbb{R}^{3 \times 3}, \mathbf{t}_i \in \mathbb{R}^3) \mid 1 \leq i \leq N\}$ as geometric features. The representation $\mathbf{Z}$ includes the RBF-based spatial distances between $C_\alpha$, $N$, $C$, and a virtual $C_\beta$ \cite{dauparas2022robust} estimated by other backbone atoms, as well as the relative sequential positions for all pairs of residues. The rigid coordinate frames \cite{jumper2021highly} are constructed using a Gram-Schmidt process from the coordinates of $C_\alpha$, $N$, and $C$, ensuring the invariance of IPA with respect to global Euclidean transformations.

\subsection*{S5. Training and sampling inference}
We present the training and sampling inference schemes for MapDiff. A formal description of the denoising training process is provided in Algorithm \ref{training_alg}. Our denoising network follows a two-step approach, where we utilize the global-aware EGNN and pre-trained IPA network to instantiate the base sequence predictor $h_\phi$ and masked sequence designer $f_\theta$, respectively. The training objective aims to minimize the cross-entropy losses $L_\mathrm{b}$ and $L_\mathrm{m}$ in the two stages, while the learnable weights $\phi$ and $\theta$ are jointly optimized through backpropagation. 

We provide the description of the sampling inference procedure in Algorithm \ref{sampling_alg}. It is a combination of DDIM with Monte-Carlo dropout to accelerate the sampling process and improve uncertainty estimation. The DDIM posterior computation is designed to skip a specific time step $k$ at each denoising sampling. Additionally, by incorporating Monte-Carlo dropout, we can compute the mean probability prediction by passing multiple stochastic forwards. Our experimental results demonstrate that this novel sampling inference substantially enhances the generative performance.

\begin{algorithm}[h]
   \caption{Sampling inference algorithm of MapDiff}
   \label{alg:training}
\begin{algorithmic}[1]
\STATE{\bf Input:} residue graph $\mathcal{G} = (\mathbf{X}, \mathbf{A}, \mathbf{E})$, where $\mathbf{X} = [\mathbf{X}^\mathrm{pos}, \mathbf{X}^\mathrm{prop}]$, pairwise representation $\mathbf{Z}$,  rigid coordinate frames $\mathcal{T}$, denoising time steps $T$, skipping step $k$, stochastic forward passes $C$
\STATE{\bf Initial:} base sequence predictor $h_\phi$, masked sequence designer $f_\theta$
\STATE \textcolor{myblue}{Sample noise from a uniform or marginal prior $\mathbf{X}_T^\mathrm{aa} \sim p(\mathbf{X}_T)$}
\STATE Activate dropout for $h_\phi$ and $f_\theta$ during inference \hfill\COMMENT{Monte-Carlo dropout}
\FOR{$c$ in $\{1,2,...,C\}$}
\FOR{$t$ in $\{T,T-k,..., 0\}$}
\STATE Predict $p_\mathrm{f}(\hat{\mathbf{X}}^\mathrm{aa}_0)$ by neural networks $h_\phi$, $f_\theta$ and $\mathbf{X}_t^\mathrm{aa}$
\STATE Compute $p_{t}^{c}(\mathbf{X}_{t - k}^\mathrm{aa} | \mathbf{X}_{t}^\mathrm{aa}, \hat{\mathbf{X}}^\mathrm{aa}_0)$ \hfill\COMMENT{DDIM posterior computation}
\ENDFOR
\ENDFOR
\STATE Compute mean prediction $p_0^\mathrm{m}(\mathbf{X}^\mathrm{aa}_{0}) = \frac{1}{C}\sum{p_{0}^{c}(\mathbf{X}_{0}^\mathrm{aa} | \mathbf{X}_{k}^\mathrm{aa}, \hat{\mathbf{X}}^\mathrm{aa}_0)}$ \hfill\COMMENT{Monte-Carlo estimation}
\STATE \bf{return} $\mathbf{X}_{0}^\mathrm{aa} \sim p_0^\mathrm{m}(\mathbf{X}^\mathrm{aa}_{0})$
\end{algorithmic}
\label{sampling_alg}
\end{algorithm}

\begin{algorithm}[htbp]
   \caption{Invariant point attention (IPA) layer}
   \label{alg:training}
\begin{algorithmic}[1]
\STATE{\bf Input:} residue representation $\mathbf{S}^{(l)} = \{\mathbf{s}_{i} \in \mathbb{R}^{d_s} \mid 1 \leq i \leq N\}$, pairwise representation $\mathbf{Z}^{(l)} = \{\mathbf{z}_{ij} \in \mathbb{R}^{d_z} \mid 1 \leq i \leq N, 1 \leq j \leq N\}$, rigid coordinate frames $\mathcal{T} = \{T_i := (\mathbf{R}_i \in \mathbb{R}^{3 \times 3}, \mathbf{t}_i \in \mathbb{R}^3) \mid 1 \leq i \leq N\}$
\STATE{\bf Parameters:} $ N_{\mathrm{head}}\ = 4, c = 32, N_{\mathrm{query \, point}} = 4, N_{\mathrm{value \, point}} = 8$

\STATE $\mathbf{q}_i^h, \mathbf{k}_i^h, \mathbf{v}_i^h = \mathrm{LinearNoBias}(\mathbf{s}_i)$ \hfill$\mathbf{q}_i^h, \mathbf{k}_i^h, \mathbf{v}_i^h \in \mathbb{R}^{d_h}, h \in \{1, \cdots, N_{\mathrm{head}}\}$
\STATE $\vec{\mathbf{q}}_i^{hp}, \vec{\mathbf{k}}_i^{hp} = \mathrm{LinearNoBias}(\mathbf{s}_i)$ \hfill$\mathbf{q}_i^{hp}, \mathbf{k}_i^{hp} \in \mathbb{R}^{3}, p \in \{1, \cdots, N_{\mathrm{query \, point}}\}$
\STATE $\vec{\mathbf{v}}_i^{hp} = \mathrm{LinearNoBias}(\mathbf{s}_i)$ \hfill$\mathbf{v}_i^{hp} \in \mathbb{R}^{3}, p \in \{1, \cdots, N_{\mathrm{value \, point}}\}$
\STATE $b_{ij}^h = \mathrm{LinearNoBias}(\mathbf{z}_{ij})$
\STATE $w_\mathrm{C} = \sqrt{\frac{2}{9 N_{\mathrm{query \, point}}}}$
\STATE $w_\mathrm{L} = \sqrt{\frac{1}{3}}$
\STATE $a_{ij}^h = \mathrm{softmax} \left(w_\mathrm{L}\left( \frac{1}{\sqrt{c}}{\mathbf{q}^h_i}^T \mathbf{k}^h_j + b_{ij}^h - \frac{\tau^h w_\mathrm{C}}{2} \sum_{p} \Vert T_i \circ \vec{\mathbf{q}}_i^{hp} - T_j \circ \vec{\mathbf{k}}_j^{hp}     \Vert^2     \right) \right)$
\STATE $\tilde{\mathbf{o}}_i^h = \sum_j a_{ij}^h \mathbf{z}_{ij}$
\STATE $\mathbf{o}_i^h = \sum_j a_{ij}^h \mathbf{v}_{j}$
\STATE $\vec{\mathbf{o}}_i^{hp} = T_i^{-1} \circ \sum_j a_{ij}^h \left( T_j \circ \vec{\mathbf{v}}_j^{hp} \right)$
\STATE $\tilde{\mathbf{s}}_i = \mathrm{Linear}\left( \mathrm{concat}_{h,p} ( \tilde{\mathbf{o}}_i^h, \mathbf{o}_i^h, \vec{\mathbf{o}}_i^{hp}, \Vert \vec{\mathbf{o}}_i^{hp} \Vert )\right)$
\STATE $\hat{\mathbf{s}}_i = \mathrm{Linear}\left( \tilde{\mathbf{s}}_i \right)$
\STATE $\tilde{\mathbf{z}}_{ij} = \mathrm{Linear}\left( \mathrm{concat}\left(\hat{\mathbf{s}}_i, \hat{\mathbf{s}}_j, \mathbf{z}_{ij} \right) \right)$

\STATE \bf{return} $\mathbf{S}^{(l+1)} = \{\tilde{\mathbf{s}}_{i} \in \mathbb{R}^{d_s} \mid 1 \leq i \leq N\}$, $\mathbf{Z}^{(l+1)} = \{\tilde{\mathbf{z}}_{ij} \in \mathbb{R}^{d_z} \mid 1 \leq i \leq N, 1 \leq j \leq N\}$
\end{algorithmic}
\label{ipa_alg}
\end{algorithm}

\subsection*{S6. Invariant point attention}
The description of an invariant point attention (IPA) layer is illustrated in Algorithm \ref{ipa_alg}. IPA plays a crucial role in AlphaFold2 \cite{jumper2021highly} as it determines the refinement process of protein structure generation. It is a geometry-aware attention mechanism that operates on a set of rigid coordinate frames and maintains SE(3)-invariance for single representation updates. We apply IPA as a masked sequence designer in MapDiff. In the original implementation, each IPA layer regards the protein as a fully connected graph and only updates the node residue representations. Following the message passing proposed by Yim et al. (2023) \cite{SE3DIFFP}, we perform an edge update by propagating the updated node information to pairwise edge representations. Finally, we extract the node representations from the last IPA layer to predict the masked residue types.

\renewcommand*{\figurename}{Supplementary Figure}
\begin{figure*}[htbp]
    \begin{center}
    \includegraphics[width=0.80\textwidth]{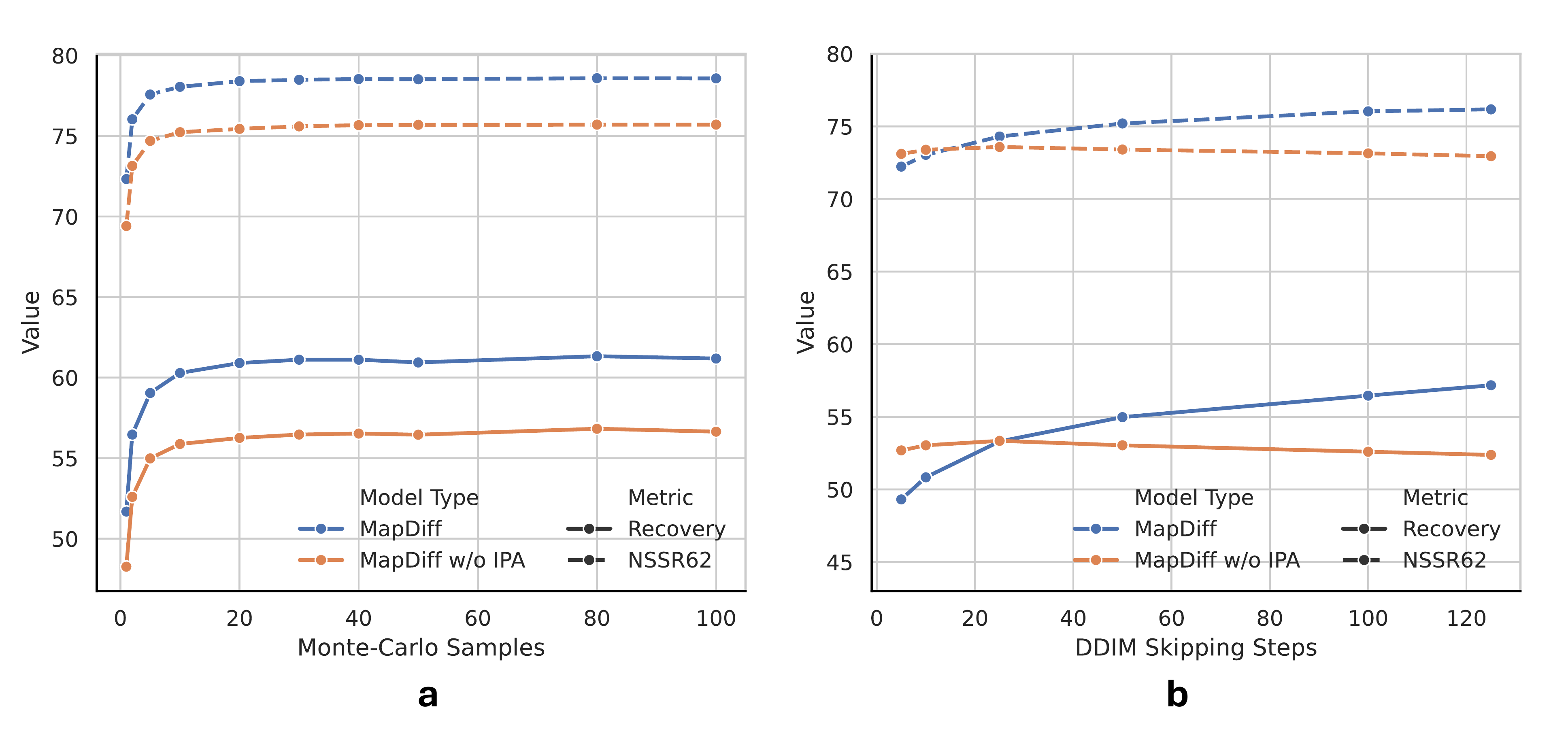}
    \end{center}
     \caption{\textbf{Sensitivity analysis of the number of Monte-Carlo samples and DDIM skipping steps.} \textbf{(a)} Median recovery and NSSR62 score of MapDiff in relation to the number of Monte Carlo samples on CATH 4.2. \textbf{(b)} Median recovery and NSSR62 score of MapDiff with respect to the number of DDIM skipping steps on CATH 4.2.}
    \label{fig:sensitivity}
\end{figure*}

\begin{figure*}[h]
    \begin{center}
    \includegraphics[width=0.82\textwidth]{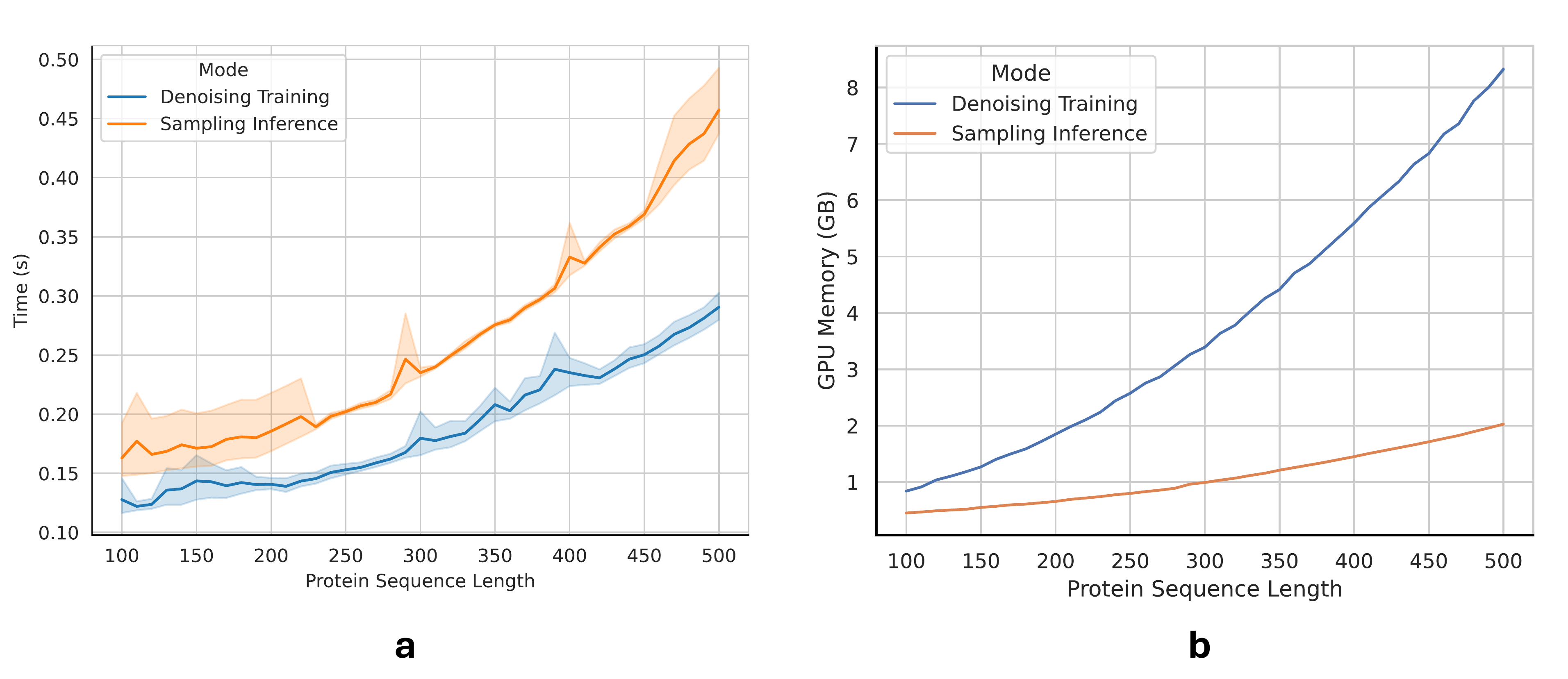}
    \end{center}
     \caption{\textbf{Runtime and GPU memory usage of MapDiff.} \textbf{(a)} Analysis of training and sampling runtimes for different protein sequence lengths, based on ten independent runs for each length. The sampling inference runtime represents the duration of one Monte-Carlo sample with DDIM (skipping steps=100) in MapDiff. The shaded area indicates the 95\% confidence interval for the runtime estimation at each protein sequence length. \textbf{(b)} Correlation between GPU memory usage and protein sequence length during training and sampling modes.}
    \label{fig:runtime}
\end{figure*}

\subsection*{S7. Sensitivity}
In Supplementary Fig. \ref{fig:sensitivity}, we analyze the performance sensitivity of MapDiff with respect to the number of Monte-Carlo samples and DDIM skipping steps. Supplementary Fig. \ref{fig:sensitivity}a demonstrates that sequence recovery and NSSR62 substantially improve with the number of included Monte Carlo samples. Notably, the model performance remains stable and high only if the number of samples reaches 20. This finding indicates that enhancing prediction uncertainty through Monte-Carlo dropout is an effective strategy for improving the diffusion-based generation process. Supplementary Fig. \ref{fig:sensitivity}b illustrates the model performance against the number of DDIM skipping steps. To isolate the performance gain from Monte-Carlo dropout, we set the number of Monte-Carlo samples to two in this validation. The results show that even as the number of skipping steps increases, MapDiff achieves higher recovery compared to the model variant without IPA. This can be attributed to the prior knowledge provided by the pre-trained IPA component, which denoises the diffusion trajectories toward more reliable sampling and, consequently, more accurate sequence generation.

\subsection*{S8. Scalability}
In Supplementary Fig. \ref{fig:runtime}, we analyze the scalability of MapDiff across denoising training and sampling inference modes. Supplementary Fig. \ref{fig:runtime}a illustrates the runtimes of model training and inference against the protein sequence length. For each protein length, we conduct ten independent runs using a single Tesla A100 GPU. We empirically observe that the runtimes of MapDiff increase almost linearly with the protein sequence length. For instance, MapDiff takes around 0.13 seconds for a short protein sequence with 100 residues during denoising training, while the runtime reaches around 0.28 seconds for a long protein sequence with 500 residues. Supplementary Fig. \ref{fig:runtime}b shows the peak GPU memory usage against the protein sequence length. Similar to the runtime, the memory usage exhibits a linear increase as the sequence length grows. During sampling inference, MapDiff only requires around 2GB of RAM for a protein sequence with 500 residues. The study highlights the scalability of MapDiff in both training and inference.

\subsection*{\textcolor{red}{S9. Evaluation metric calculation}}
\textcolor{red}{We evaluated the accuracy of generated sequences using recovery rate, native sequence similarity recovery (NSSR) and perplexity, while evaluating the quality of refolded structures with root mean square deviation (RMSD), global distance test-total score (GDT-TS), template modelling score (TM-score), predicted local distance difference test (pLDDT), predicted aligned eerro (pAE), and predicted template modelling (PTM).}

\textcolor{red}{\textbf{Recovery rate} indicates the proportion of accurately predicted amino acids in a protein sequence, which is calculated as:}
    \begin{align}
        \textcolor{red}{\mathrm{Recovery}\ \mathrm{Rate} = \frac{1}{N} \sum_{i=1}^{N} \mathbbm{1} (\hat{s}_i = s_i)}
    \end{align}
\noindent \textcolor{red}{where $N$ is the length of a protein sequence, $\hat{s}_i$ is the predicted amino acid at position $i$, $\mathbbm{1} (\hat{s}_i = s_i)$ is an indicator function that equals 1 if the predicted amino acid matches the true amino acid, and 0 otherwise.}

\textcolor{red}{\textbf{NSSR} evaluates the similarity between the predicted and native residues via the Blocks Substitution Matrix (BLOSUM), where each residue pair contributes to a positive prediction if their BLOSUM score is greater than zero. It is calculated as:}
    \begin{align}
        \textcolor{red}{\mathrm{NSSR} = \frac{1}{N} \sum_{i=1}^{N} \mathbbm{1} \big( B(\hat{s}_i, s_i) > 0 \big),}
    \end{align}
\noindent \textcolor{red}{where $B(\hat{s}_i, s_i) > 0$ is the BLOSUM similarity score for the predicted and true residue pair $(s_i, \hat{s}_i)$ at position $i$.}
    
\textcolor{red}{\textbf{Perplexity} measures the alignment between the predicted amino acid probabilities and the true amino acid types at each residue position, which is calculated as follows:}
    \begin{align}
    \textcolor{red}{\mathrm{Perplexity} = \exp \left( -\frac{1}{M} \sum_{i=1}^{M} \log P(s_i | X) \right),}
    \end{align}
\noindent \textcolor{red}{where $M$ is the number of residues, and $P(s_i | X)$ is the probability predicted by the model to the true amino acid $s_{i}$ at position $i$, given the protein structure $X$.}

\textcolor{red}{\textbf{RMSD}, \textbf{TM-score and} \textbf{GDT-TS} are three common metrics to measure the discrepancies between a predicted protein structure and its native counterpart, which are calculated as follows:}
    \begin{align}
     \textcolor{red}{\mathrm{RMSD} = \sqrt{\frac{1}{N} \sum_{i=1}^{N} \|\mathbf{r}_i^{\text{pred}} - \mathbf{r}_i^{\text{true}}\|^2},}
    \end{align}
\noindent \textcolor{red}{where $\mathbf{r}_i^{\text{pred}}$ and $\mathbf{r}_i^{\text{true}}$ denote the predicted and true coordinates of the $C_{\alpha}$ atom at residue $i$, respectively.}

    \begin{align}
        \textcolor{red}{\mathrm{TM\text{-}score} = \frac{1}{N} \sum_{i=1}^{L} \frac{1}{1 + \left( \frac{d_i}{d_0(L)} \right)^2},}
    \end{align}
    \textcolor{red}{where $L$ is the number of aligned residues, $d_i$ represents the $C_\alpha$ distance between the predicted and native structure at residue $i$, and $d_0^{L}$ is a length-dependent normalization factor.}

    \begin{align}
        \textcolor{red}{\mathrm{GDT\text{-}TS} = \frac{1}{4} \sum_{d \in \{1, 2, 4, 8\} \text{Å}} \frac{N_d}{N} \times 100,}
    \end{align}
\noindent \textcolor{red}{where $N_d$ is the number of residues whose $C_\alpha$ distance in the predicted structure are within $d$ Å of the native structure.}

\textcolor{red}{\textbf{pLDDT, pAE and PTM} are three confidence metrics in AlphaFold2 that estimate the reliability of predicted structures without relying on native structures during inference. Instead, AlphaFold2 predicts their distance probability distributions via deep learning. pLDDT is a per-residue confidence score ranging from 0 to 100, where higher values indicate greater local accuracy. pAE measures the expected positional error between residue pairs, helping assess domain flexibility and global alignment confidence. PTM estimates the global accuracy of the predicted structures, reflecting how well the overall structure is likely to be correct.}


\end{document}